\documentclass[aps,prd,twocolumn,showpacs,superscriptaddress,groupedaddress,amsmath,amssymb]{revtex4-1}

\usepackage{graphicx}
\usepackage{dcolumn}
\usepackage{bm}
\usepackage{hyperref}
\usepackage{multirow}

\usepackage{verbatim}
\usepackage{longtable}

\def\be{\begin{equation}}
\def\bea{\begin{eqnarray}}
\def\eea{\end{eqnarray}}
\def\ee{\end{equation}}
\def\bi{\begin{itemize}}
\def\ei{\end{itemize}}

\def\bn{\begin{enumerate}}
\def\en{\end{enumerate}}

\def\be{\begin{equation}}
\def\ee{\end{equation}}
\def\bea{\begin{eqnarray}}
\def\eea{\end{eqnarray}}

\def\beq{\begin{equation}}
\def\eeq{\end{equation}}

\begin{document}

\title{First low frequency all-sky search for continuous gravitational wave signals}

\author{J.~Aasi,$^{1}$
B.~P.~Abbott,$^{1}$
R.~Abbott,$^{1}$
T.~D.~Abbott,$^{2}$
M.~R.~Abernathy,$^{1}$
F.~Acernese,$^{3,4}$
K.~Ackley,$^{5}$
C.~Adams,$^{6}$
T.~Adams,$^{7,8}$
P.~Addesso,$^{9}$
% ...
}

\affiliation {$^{1}$LIGO---California Institute of Technology, Pasadena, CA 91125, USA }
\affiliation {$^{2}$Louisiana State University, Baton Rouge, LA 70803, USA }
\affiliation {$^{3}$Universit\`a di Salerno, Fisciano, I-84084 Salerno, Italy }
\affiliation {$^{4}$INFN, Sezione di Napoli, Complesso Universitario di Monte S.Angelo, I-80126 Napoli, Italy }
\affiliation {$^{5}$University of Florida, Gainesville, FL 32611, USA }
\affiliation {$^{6}$LIGO Livingston Observatory, Livingston, LA 70754, USA }
\affiliation {$^{7}$Cardiff University, Cardiff CF24 3AA, United Kingdom }
\affiliation {$^{8}$Laboratoire d'Annecy-le-Vieux de Physique des Particules (LAPP), Universit\'e Savoie Mont Blanc, CNRS/IN2P3, F-74941 Annecy-le-Vieux, France }
\affiliation {$^{9}$University of Sannio at Benevento, I-82100 Benevento, Italy and INFN, Sezione di Napoli, I-80100 Napoli, Italy }
% ...


\begin{abstract} 
  In this paper we present the results of the first low frequency all-sky search of continuous gravitational wave signals conducted on Virgo VSR2 and VSR4 data. The search covered the full sky, a frequency range between 20 Hz and 128 Hz with a range of spin-down between
  $-1.0 \times 10^{-10}$ Hz/s and $+1.5 \times 10^{-11}$ Hz/s, and was based on a hierarchical approach. The starting point was a set of short Fast Fourier Transforms (FFT), of length 8192 seconds, built from the calibrated strain  data. 
  Aggressive data cleaning, both in the time and frequency domains, has been done in order to remove, as much as possible, the effect of disturbances of instrumental origin.
  On each dataset a number of candidates has been selected, using the FrequencyHough transform in an incoherent step. Only coincident candidates among VSR2 and VSR4 have been examined in order to strongly reduce the false alarm
  probability, and the most significant candidates have been selected. The criteria we have used for candidate selection and for the coincidence step greatly reduce the harmful effect of large instrumental artifacts.   
  Selected candidates have been subject to a follow-up by constructing a new set of longer FFTs followed by a further incoherent analysis, still based on the FrequencyHough transform. No evidence for continuous gravitational wave signals was found, therefore we have set a population-based joint
  VSR2-VSR4 90$\%$ confidence level upper limit on the dimensionless gravitational wave strain in the frequency range between 20 Hz and 128 Hz.
This is the first all-sky search for continuous gravitational waves conducted, on data of ground-based interferometric detectors, at frequencies below 50 Hz. We set upper limits in the range between about $10^{-24}$ and $2\times 10^{-23}$ at most frequencies. Our upper limits on signal strain show an improvement of up to a factor of $\sim$2 with respect to the results of previous all-sky searches at frequencies below $80~\mathrm{Hz}$. 
\end{abstract}

\pacs{04.80Nn,95.55Ym, 97.60Gb, 07.05Kf}

\maketitle

\section{Introduction}
\label{sec:intro}
Continuous gravitational wave signals (CW) emitted by asymmetric spinning neutron stars are among the sources currently sought in the data of interferometric gravitational wave detectors.
The search for signals emitted by spinning neutron stars with no electromagnetic counterpart requires the exploration of a large portion of the source parameter space, consisting of the source position, 
signal frequency and signal frequency time-derivative (spin-down). This kind of search, called all-sky, cannot be based on fully coherent methods, as in targeted searches for known pulsars, see e.g. \cite{ref:vela_vsr2,ref:vsr4_kp}, because of the 
huge computational resources that would be required. 

For this reason various hierachical analysis pipelines, based on the alternation of coherent and incoherent steps, have been developed \cite{ref:powerflux}, 
\cite{ref:ein@ho}, \cite{ref:krolsky}, \cite{ref:alisky}, \cite{ref:freqhoughmethod}. They allow us to dramatically reduce the computational burden of the analysis, at the cost of a small sensitivity loss. In this paper we present the 
results of the first 
all-sky search for CW signals using the data of Virgo science runs VSR2 and VSR4 (discussed in Sec. \ref{sec:ITF}). The analysis has been carried out on the frequency band 20-128 Hz, using an efficient hierarchical analysis pipeline, based on the FrequencyHough transform \cite{ref:freqhoughmethod}. No detection was made, so we established upper limits on signal strain amplitude as a function of the frequency. 
Frequencies below 50 Hz have never been considered in all-sky searches for CW signals, and the estimated joint sensitivity of Virgo VSR2 and VSR4 data is better than that of data from LIGO science runs S5 and S6 below about 60-70 Hz. Moreover, lower frequencies could potentially offer promising sources. Higher frequency signals would be in principle easier to detect because of their high signal amplitudes at fixed distance and ellipticity (see Eq. \ref{eq:h0}). On the other hand, neutron stars with no electromagnetic counterpart, which are the main target of an all-sky search, could have a spin rate distribution significantly different with respect to standard pulsars. Then, we cannot exclude that a substantial fraction of neutron stars emits gravitational waves with frequency in the range between 20 Hz and about 100 Hz. 
This is particularly true when considering young, unrecycled, neutron stars, which could be more distorted than older objects. Below about 20 Hz the detector sensitivity significantly worsens and the noise is highly non-stationary making the analysis pointless. 
A search at low frequency, as described below, could detect signals from a potentially significant population of nearby neutron stars.

The plan of the paper is as follows. In Sec. \ref{sec:signal} we describe the kind of gravitational wave (GW) signal we are searching for. In Sec. \ref{sec:ITF} we discuss the Virgo detector performance during VSR2 and VSR4 runs. In Sec. \ref{sec:proc} we briefly recap the analysis procedure, referring the reader to \cite{ref:freqhoughmethod} for more details. Section \ref{sec:clean} is focused on the cleaning steps applied at different stages of the analysis. Section \ref{sec:cand} is dedicated to candidate selection, and Sec. \ref{sec:clusters} to their clustering and coincidences. Section \ref{sec:followup} deals with the follow-up of candidates surviving the coincidence step.
Section \ref{sec:HI} is dedicated to validation tests of the analysis pipeline, by using hardware-injected signals in VSR2 and VSR4 data. 
In Sec. \ref{sec:ul} a joint upper limit on signal strain amplitude is derived as a function of the search frequency. 
Conclusions and future prospects are presented in Sec. \ref{sec:concl}. Appendix \ref{sec:finalcandlist} contains a list of the 108 candidates for which the follow-up has been done, along with their main parameters. Appendix \ref{sec:outl} is devoted to a deeper analysis of the three outliers found. Appendix \ref{sec:excluded} contains the list of frequency intervals excluded from the computation of the upper limits.     

\section{The Signal}
\label{sec:signal}
The expected quadrupolar GW signal from a non-axisymmetric neutron star steadily spinning around one of its principal axes has a frequency $f_0$ 
twice the rotation frequency $f_\mathrm{rot}$, with a strain at the detector of \cite{ref:fivevect,ref:freqhoughmethod}
\be
h(t)=H_0(H_+A^++H_{\times}A^{\times})e^{\jmath \left(\omega(t)t+\Phi_0\right)},
\label{eq:hoft}
\ee
where taking the real part is understood and where $\Phi_0$ is an initial phase. The signal's time-dependent angular frequency $\omega(t)$ will be discussed below. The two complex amplitudes $H_+$ and $H_{\times}$ are given respectively by
\begin{equation}
H_+=\frac{\cos{2\psi}-\jmath \eta \sin{2\psi}}{\sqrt{1+\eta^2}},
\label{eq:Hp}
\end{equation}
\begin{equation}
H_{\times}=\frac{\sin{2\psi}+\jmath \eta \cos{2\psi}}{\sqrt{1+\eta^2}},
\label{eq:Hc}
\end{equation}
in which $\eta$ is the ratio of the polarization ellipse semi-minor to semi-major axis, and the polarization angle $\psi$ defines the direction of the major axis with respect to the celestial parallel of the source (counterclockwise). The parameter $\eta$ varies in the range $[-1,1]$, where $\eta=0$ for a linearly polarized wave, while $\eta=\pm 1$ for a circularly polarized wave ($\eta=1$ if the circular rotation is counterclockwise). 
The functions $A^{+,\times}$ describe  the detector response as a function of time, with a periodicity of one and two sidereal periods, and depend on the source position, detector position and orientation on the Earth \cite{ref:fivevect}. 

As discussed in \cite{ref:vela_vsr2}, the strain described by Eq.(\ref{eq:hoft}) is equivalent to the standard expression (see e.g. \cite{ref:jks})
\begin{align}
h(t) =  &\frac{1}{2}F_+(t, \psi)h_0(1+\cos{}^2\iota)\cos{\Phi(t)} \\ \nonumber 
& + F_{\times}(t, \psi)h_0\cos{\iota}\sin{\Phi(t)}
\label{eq:hclass}
\end{align}
Here $F_+,~F_{\times}$ are the standard beam-pattern functions and $\iota$ is the angle between the star's rotation axis and the line of sight. The amplitude parameter
\begin{equation}\label{eq:h0}
h_0=\frac{4\pi^2G}{c^4}\frac{I_{zz}\varepsilon f^2_0}{d}
\end{equation}
depends on the signal frequency $f_0$ and on the source distance $d$; on $I_{zz}$, the star's moment of inertia with respect to the 
principal axis aligned with the rotation axis; and on $\varepsilon$, which is the fiducial equatorial ellipticity expressed in terms of principal moments of inertia as
\begin{equation}
\varepsilon=\frac{I_{xx}-I_{yy}}{I_{zz}}.
\end{equation}
It must be stressed that it is not the fiducial ellipticity but the quadrupole moment $Q_{22}\propto I_{zz}\varepsilon$ that, in case of detection, can be measured independently of any assumption about the star's equation of state and moment of inertia (assuming the source distance can be also estimated).
There exist estimates of the maximum ellipticity a neutron star can sustain from both elastic and magnetic deformations. In the elastic case, these maxima depend strongly on the breaking strain of the solid portion sustaining the deformation (see, e.g., \cite{ref:horo}, \cite{ref:hoff} for calculations for the crust) as well as on the star's structure and equation of state, and the possible presence of exotic phases in the star’s interior (like in hybrid or strange quark stars, see, e.g., \cite{ref:john_owen}). In the magnetic case, the deformation depends on the strength and configuration of the star's internal magnetic field (see, e.g., \cite{ref:ciolfi}). However, the actual ellipticity of a given neutron star is unknown---the best we have are observational upper limits.
The relations between $H_0,~\eta$ and $h_0,~\iota$ are given, e.g., in \cite{ref:fivevect}. 
In  Eq. \ref{eq:hoft} the signal angular frequency $\omega(t)$ is a function of time, therefore the signal phase 
\begin{equation}
\Phi(t)=\int_{t_0}^t \omega(t')dt'
\label{eq:phit}
\end{equation} 
is not that of a simple monochromatic signal. It depends on the rotational frequency and frequency derivatives of 
the neutron star, as well as on Doppler and propagation effects. In particular,
the received Doppler-shifted frequency $f(t)$ is related to the emitted frequency $f_0(t)$ by the well-known relation (valid in the non-relativistic approximation)
\begin{equation}
f(t)=\frac{1}{2\pi}\frac{d\Phi(t)}{dt}= f_0(t) \left(1+\frac{\vec{v}(t)
\cdot \hat{n}}{c}\right),
\label{eq:fdopp}
\end{equation}
where $\vec{v}$ is the detector velocity with respect to the Solar System barycenter (SSB), $\hat{n}$ is the unit vector in the direction to the source from the SSB and $c$ is the light speed. A smaller 
relativistic effect, namely the {\it Einstein delay}, is not relevant for the incoherent step of the search described in Sec. \ref{sec:proc}, due to the use of short length FFTs, and has been therfore neglected. On the contrary, it has been taken into account in the candidate follow-up, described in Sec. \ref{sec:followup}, specifically when a coherent analysis using candidate parameters is done.

The intrinsic signal frequency $f_0(t)$ slowly decreases in time due to the source's spin-down, associated with the rotational energy loss following emission of electromagnetic and/or gravitational radiation. The spin-down can be described through a series expansion
\begin{equation}
f_0(t)=f_0+\dot{f}_0(t-t_0)+\frac{\ddot{f}_0}{2}(t-t_0)^2+...
\label{eq:sd}
\end{equation}
In general, the frequency evolution of a CW depends on 3+$s$ parameters: position, frequency and $s$ spin-down parameters. In the all-sky search described in this paper we need to take into account only the first spin-down ($s=1$) parameter (see Sec. \ref{sec:proc}).

\section{Instrumental performance during VSR2 and VSR4 runs}
\label{sec:ITF}
Interferometric GW detectors, such as LIGO \cite{ref:ligo}, Virgo \cite{ref:virgo}, and GEO \cite{ref:geo}, have collected years of data, from 2002 to 2011.
For the analysis described in this paper we have used calibrated data from the Virgo VSR2 and VSR4 science runs. The VSR3 run was characterized by a diminished sensitivity level and poor data quality (highly non-stationary data, large glitch rate), and so was not included in this analysis.
The VSR2 run began on July
7th, 2009 (21:00 UTC) and ended on January 8th, 2010 (22:00 UTC). The duty cycle was $80.4\%$, resulting
in a total of $\sim 149$ days of {\it science mode} data, divided among $361$ segments. The data used in the analysis have been produced using the most up-to-date calibration parameters and reconstruction procedure. The associated
systematic error amounts to $5.5\%$ in amplitude and $\sim 50$~mrad in
phase \cite{ref:virgovsr2performance}.

\begin{figure*}[!htbp]
\includegraphics[width=12cm]{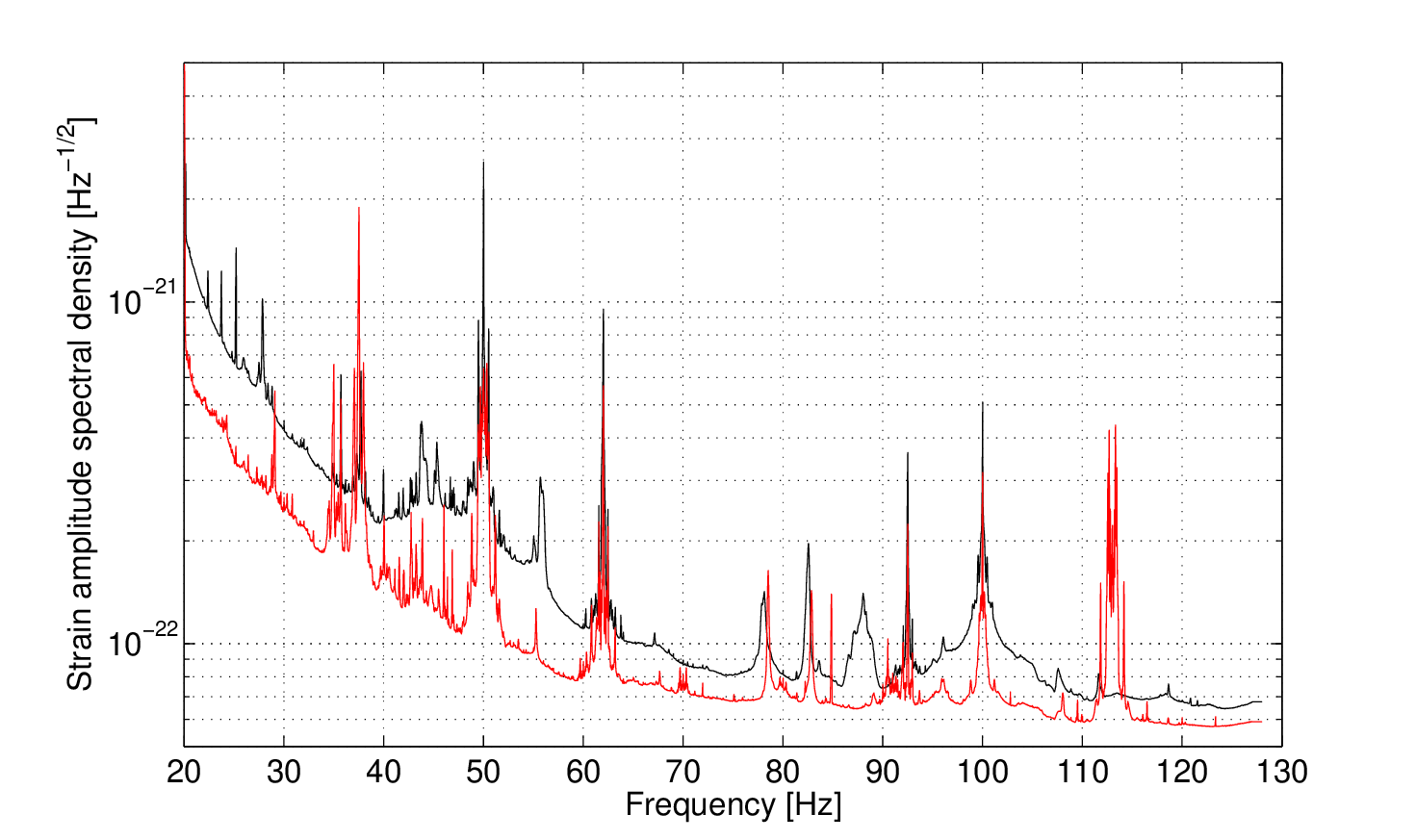} 
\caption{VSR2 (darker, black in the color version) and VSR4 (lighter, red in the color version) average strain amplitude spectral density in the frequency range from 20 Hz up to 128 Hz.}
\label{fig:experimental}
\end{figure*}

The VSR4 run extended from June 3rd, 2011 (10:27 UTC) to September 5th, 2011 (13:26 UTC), with a duty factor of about 81$\%$, corresponding to an effective duration of 76 days. Calibration 
uncertainties amounted to 7.5$\%$ in amplitude and $(40+50f_{\rm kHz})$ mrad in phase up to 500 Hz, where $f_{\rm kHz}$ is the frequency in kilohertz \cite{ref:vsr4cal}. The uncertainty on the 
amplitude contributes to the uncertainty on the upper limit on signal amplitude, together with that coming from the finite size of the Monte Carlo simulation used to compute it (see Sec. \ref{sec:ul}). A calibration error on the phase of this size can be shown to have a negligible impact on the analysis \cite{ref:vela_vsr2}. The low-frequency sensitivity of VSR4 was significantly better, up to a factor of 2, than that of previous Virgo runs, primarily due to the use of monolithic mirror suspensions, and nearly in agreement with the design sensitivity of the initial Virgo interferometer \cite{ref:virgo}. This represents a remarkable improvement considering that, being the gravitational wave strain proportional to the inverse of the distance to the source, a factor of 2 in sensitivity corresponds to an increase of a factor of 8 in the accessible volume of space (assuming a homogeneous source distribution). 
We show in  Fig. \ref{fig:experimental} the average experimental strain amplitude spectral density for VSR2 and VSR4, in the frequency range 20-128 Hz, obtained by making an average of the periodograms (squared modulus of the FFTs) stored in the short FFT database (see next section).

\section{The analysis procedure}
\label{sec:proc}

All-sky searches are intractable using completely coherent methods because of the huge size of the
parameter space, which poses challenging computational problems \cite{ref:stack1}, \cite{sergioqui}. Moreover, a completely coherent 
search would not be robust against unpredictable phase variations of the signal 
during the observation time. 

For these reasons hierarchical schemes have been developed.
The hierarchical scheme we have used for this analysis has been described in detail in \cite{ref:freqhoughmethod}. In this section we briefly recall the main steps. The analysis starts from the detector calibrated data, sampled at 4096 Hz. The first step consists of constructing a database of {\it short Discrete Fourier Transforms} (SFDB) \cite{SFDB}, computed through the FFT algorithm (the FFT is just an efficient algorithm to compute Discrete Fourier Transforms (DFT), but for historical reasons and for consistency with previous papers we will use the term FFT instead of DFT). Each FFT covers the frequency range from $20$ to $128\mathrm{Hz}$ and is built from a data chunk of duration ({\it coherence time}) short enough such that if a signal is present, its frequency (modified by Doppler and
spin-down) does not shift more than a frequency bin. The FFT duration for this search is 8192 seconds. This corresponds to a natural frequency resolution $\delta f=1.22 \times 10^{-4}$ Hz. The FFTs are interlaced by half and windowed with a Tukey window with a width parameter $\alpha=0.5$ \cite{ref:tuckey}. Before constructing the SFDB, short strong time domain disturbances are removed from the data. This is the first of several cleaning steps applied to the data (see Sec. \ref{sec:clean}). The total number of FFTs for VSR2 run is 3896 and for VSR4 run is 1978. 

\begin{figure*}[!htbp]
\includegraphics[width=12cm]{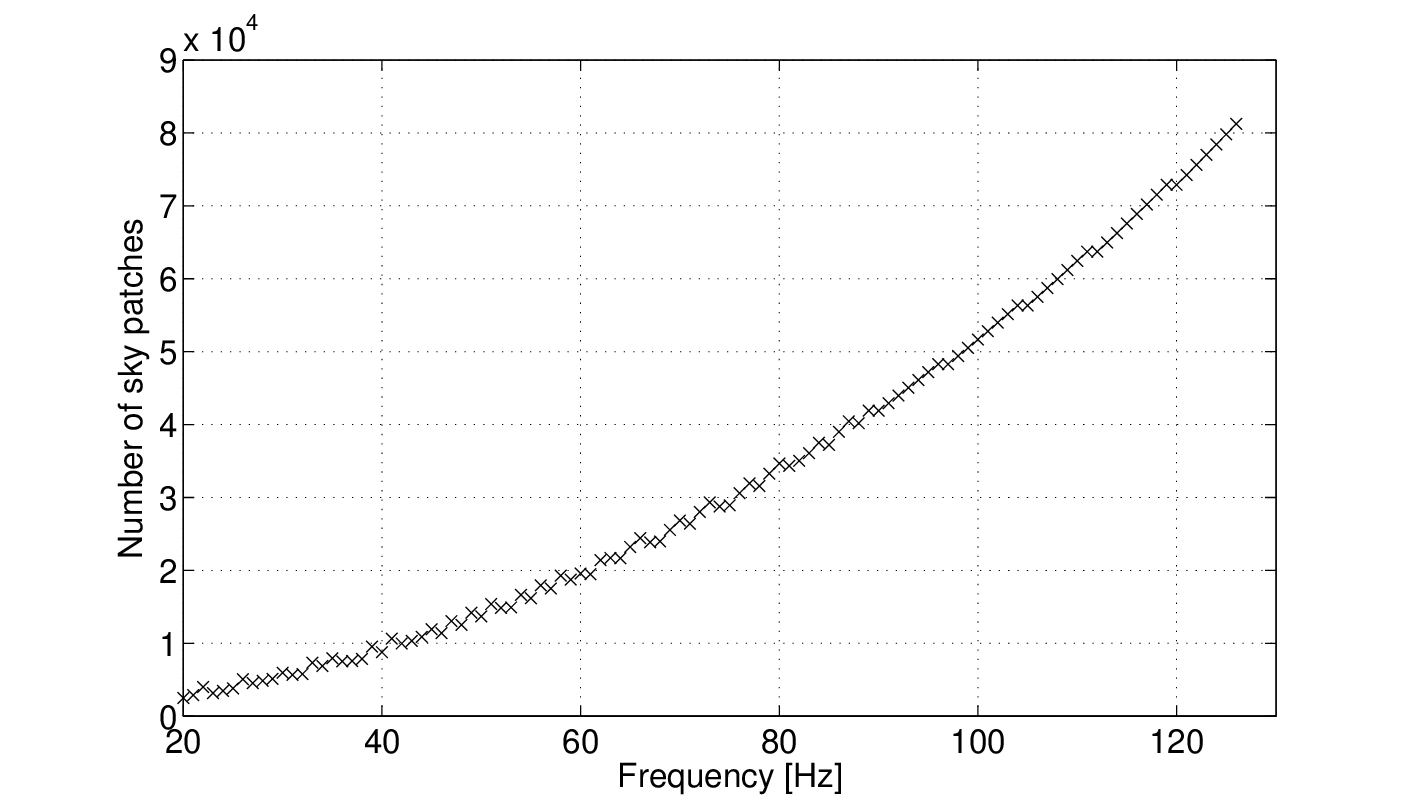}
\caption{Number of sky patches in every 1 Hz band, from 20 Hz up to 128 Hz. The frequency on abscissa axis indicates the beginning frequency of each band. The number of patches increases with the square of the frequency, and is fixed by the highest frequency of each 1 Hz band.}  
\label{fig:npatchesperband}
\end{figure*}

From the SFDB we create a time-frequency map, called the {\it peakmap} \cite{sec:Cleaning}. This is obtained by selecting the
most significant local maxima (which we call {\it peaks}) of the square root of equalized periodograms, obtained by dividing the periodogram by an auto-regressive average spectrum estimation. The threshold for peak selection has been
chosen equal to $\sqrt{2.5}=1.58$ \cite{ref:freqhoughmethod} which, in the ideal case of Gaussian noise, would correspond to a probability of selecting a peak of 0.0755. The peakmap is cleaned by removing peaks
clearly due to disturbances, as explained in Sec. \ref{sec:clean}. 
The peakmap is then corrected for the Doppler shift for the different sky directions, by shifting the frequency of the peaks by an amount corresponding to the variation the frequency of a signal
coming from a given direction would be subject to at a given time.   
A {\it coarse} grid in the sky is used in this stage of the analysis.
The grid is built using ecliptic coordinates, as described in \cite{ref:freqhoughmethod}. Figure \ref{fig:npatchesperband} shows the number of sky points (``patches'') as a function of the frequency (in steps of 1 Hz), for both VSR2 and VSR4 analyses. The number of patches increases with the square of the frequency, and ranges from 2492 at 20 Hz to 81244 at 128 Hz. The total number of sky patches is $N_{\mathrm{sky}}$ $\approx 3.5 \times 10^6$. 

Each corrected peakmap is the input of the incoherent step, based on the FrequencyHough transform \cite{HoughFFdot,ref:freqhoughmethod}. This is a very efficient implementation of the Hough transform (see \cite{HoughFFdot} for efficiency tests and comparison with a different implementation) which, for every sky position, maps the points of the peakmap into the signal frequency/spin-down plane. In the FrequencyHough transform we take into account slowly varying  non-stationarity in the noise and the varying detector sensitivity caused by
the time-dependent radiation pattern \cite{AdaHou}, \cite{AdaHou2}. 
Furthermore, the frequency/spin-down plane is discretized by building a suitable grid \cite{ref:freqhoughmethod}. As the transformation from the peakmap to the Hough plane is not computationally bounded 
by the size of the frequency bin (which only affects the size of the Hough map) we have increased the frequency resolution by a factor of 10, with respect to the natural step size $\delta f$, in order to reduce the digitalization loss, so that the actual resolution is $\delta f_H = \delta f/10 = 1.22 \times 10^{-5}$ Hz. 

We have searched approximately over the spin-down range $[-1.0 \times 10^{-10}, +1.5 \times 10^{-11}]$ Hz/s. This choice has been dictated by the need to not increase too much the computational load of the analysis while, at the same time, covering a range of spin-down values including the values measured for most known pulsars. Given the spin-down bin width scales as $T_{\mathrm{obs}}^{-1}$ ($\delta \dot{f} = \delta f/T_{\mathrm{obs}}$), this implies a different number of spin-down values for the two data sets: $N_{\mathrm{sd}}$=16  for VSR2 with a resolution of $\delta \dot{f} = 7.63 \times 10^{-12}$ Hz/s, and $N_{\mathrm{sd}}$=9  for VSR4 with a resolution of $\delta \dot{f}  = 1.5 \times 10^{-11}$ Hz/s. The corresponding minimum gravitational-wave spin-down age,
defined as $\tau_{\mathrm{min}}(f)=f/4N_{\mathrm{sd}}\delta \dot{f}$ (where $4N_{\mathrm{sd}}\delta \dot{f}$ is the absolute value of the maximum spin-down we have searched over), is a function of the frequency, going from $\sim$1600 yr to $\sim$10200 yr for VSR2 and from $\sim$1500 yr to $\sim$9700 yr for VSR4. These values are large enough that only the first order spin-down is needed in the analysis \cite{ref:freqhoughmethod}. 
In Tab. \ref{tab:par} some quantities referring to the FFTs and peakmaps of VSR2 and VSR4 data sets are given. Tab. \ref{tab:values} contains a summary of the main parameters of the coarse step, among which the exact spin-down range considered for VSR2 and VSR4 analyses.
\begin{table}[!htbp]
\begin{center}
\begin{tabular}{|c|c|c|c|c|c|}
\hline
 Run   & $T_{\mathrm{obs}}$ & $T_{\mathrm{start}}$  & $T_{\mathrm{FFT}} $ & $ N_{\mathrm{peaks}}$  \\
  $$  &  [days] &  [mjd] & [s] & (after vetoes)   \\
\hline
VSR2  & 185 & 55112 & $8192$ & $191771835 $  \\
\hline
VSR4  & 95 & 55762 & $8192$ & $93896752 $  \\
\hline
\end{tabular}
\caption{Some quantities referring to FFTs and peakmaps. For each run $T_{\mathrm{obs}}$ is the run duration, $T_{\mathrm{start}}$ is the run start epoch, $T_{\mathrm{FFT}}$ is the FFT time length, and $N_{\mathrm{peaks}}$ is the number of peaks in the peakmap, after applying all the vetoing procedures.}
\label{tab:par}
\end{center}
\end{table}
\begin{table*}[!htbp]
\begin{center}
\begin{tabular}{|c|c|c|c|c|c|c|c|}
\hline
 Run   & $\delta f_H $ & $N_{\mathrm{f}}$ &$\delta \dot{f}$  & $N_{\mathrm{sd}}$ & $\Delta \dot{f}$ & $\tau_{\mathrm{min}} $ & $N_{\mathrm{sky}}$   \\
   &  [Hz] &  &[Hz/s] &  & [Hz/s]  & [years]  & \\
\hline
 VSR2  & $ 1.22\cdot 10^{-5} $&8,847,360 &$7.63\cdot 10^{-12} $  & $16$ & [-9.91, 1.52]$\cdot 10^{-11}$ & 1600-10200 & 3,528,767 \\
\hline
 VSR4  & $ 1.22\cdot 10^{-5} $ &8,847,360 &$1.50\cdot 10^{-11}$  & $9$ & [-10.5 , 1.50]$\cdot 10^{-11}$ & 1500-9700 & 3,528,767 \\
\hline
\end{tabular}
\caption{Summary of the main parameters for the coarse step of the analysis. $\delta f_H$ is the frequency bin, $N_{\mathrm{f}}$ is the number of frequency bins in the analyzed band, $\delta \dot{f}$ is the spin-down bin, $N_{\mathrm{sd}}$ is the number of spin-down steps, $\Delta \dot{f}$ is the range of spin-down covered in the analysis, $\tau_{\mathrm{min}}$ is the corresponding minimum spin-down age, $N_{\mathrm{sky}}$ is the total number of sky patches.}
\label{tab:values}
\end{center}
\end{table*}

For a given sky position, the result of the FrequencyHough transform is a histogram in the signal frequency/spin-down plane, called the Hough map. The most significant candidates, i.e. the bins of the Hough map with the highest amplitude, are then selected using an effective way to avoid being blinded by particularly disturbed frequency bands (see Sec. \ref{sec:cand} and \cite{ref:freqhoughmethod}). 
For each coarse (or {\it raw}) candidate a refined search, still based on the FrequencyHough, is run again on the 
neighborhood of the candidate parameters and the final {\it first-level} refined candidates are selected. 
The refinement results in a reduction of the digitalization effects, that is the sensitivity loss due to the use of a discrete grid in the parameter space. With regard to the frequency, which was already refined at the coarse step, no further 
over-resolution occurs. For the spin-down we have used an over-resolution factor $K_{\dot{f}}$ =6, {\it i.e.}, the coarse interval between the spin-down of each candidate and the next value (on both sides) is divided in 6 pieces. The refined search range includes  $2\, K_{\dot{f}}$=12 bins on the left of the coarse original value, and $(2\, K_{\dot{f}}-1)$=11 bins on the right, so that two coarse bins are covered on both sides. This choice is dictated by the fact that the refinement is also done in parallel for the position of the source and, because of parameter correlation, 
a coarse candidate could be found with a refined spin-down value outside the original coarse bin.
Using an over-resolution factor $K_{\dot{f}}$ =6 is a compromise between the reduction of digitization effects and the increase of computational load.  

The refinement in the sky position for every candidate is performed by using a 
rectangular region centered at the candidate coordinates. The distance between the estimated latitude (longitude) and the next 
latitude (longitude) point in the coarse grid is divided into  ${K}_{\mathrm{sky}}$= 5  points, so that 25 sky points are considered in total, see discussion in Sec. IX-C of \cite{ref:freqhoughmethod}. 

From a practical point of view, the incoherent step of the analysis has been done by splitting the full parameter space to be explored in several independent jobs. Each job covered a frequency band of 5 Hz, the full spin-down range, and a portion of the sky (the extent of which depended on the frequency band and was chosen to maintain balanced job durations). The full set of jobs was run on the European Grid Infrastructure (http://www.egi.eu/). Overall, about 7000 jobs were run, with a total computational load of about 22,000 CPU$\cdot$hours.  

Candidates found in the analysis of VSR2 and VSR4 data are then {\it clustered}, grouping together those occupying nearby points in the parameter space. This is done to improve the computational efficiency of the next steps of the analysis. In order to significantly reduce the false alarm
probability, {\it coincidences} are required among clusters of candidates obtained from the the two data sets. 
The most significant coincident candidates are subject to a {\it follow-up} with greater coherence time, in order to confirm or discard them. Candidate selection and analysis are described in some detail in Sec. \ref{sec:cand}-\ref{sec:followup}.

\section{Data cleaning}
\label{sec:clean}
Time and frequency domain disturbances in detector data affect
the search and, if not properly removed, can significantly degrade the search sensitivity, in the worst case blinding the search at certain times or in certain frequency bands.
The effects can vary depending on the nature and amplitude of the disturbance.  
As described in \cite{ref:freqhoughmethod}, we apply cleaning procedures to safely 
remove such disturbances or reduce their effect, without contaminating a possible CW signal.
The disturbances can be catalogued as ``time domain glitches'', which enhance
the noise level of the detector in a wide frequency band; 
``spectral lines of constant frequency'', in most cases of known origin, like calibration lines or lines whose
origin has been discovered by studying the behaviour of the detector and the surrounding environment, or ``spectral wandering lines'', where the frequency of the disturbance 
changes in time (often of unknown origin and present only for a few days or even hours).
Time-domain glitches are removed during the construction of the SFDB. 

\begin{figure*}[!htbp]
\includegraphics[width=8cm]{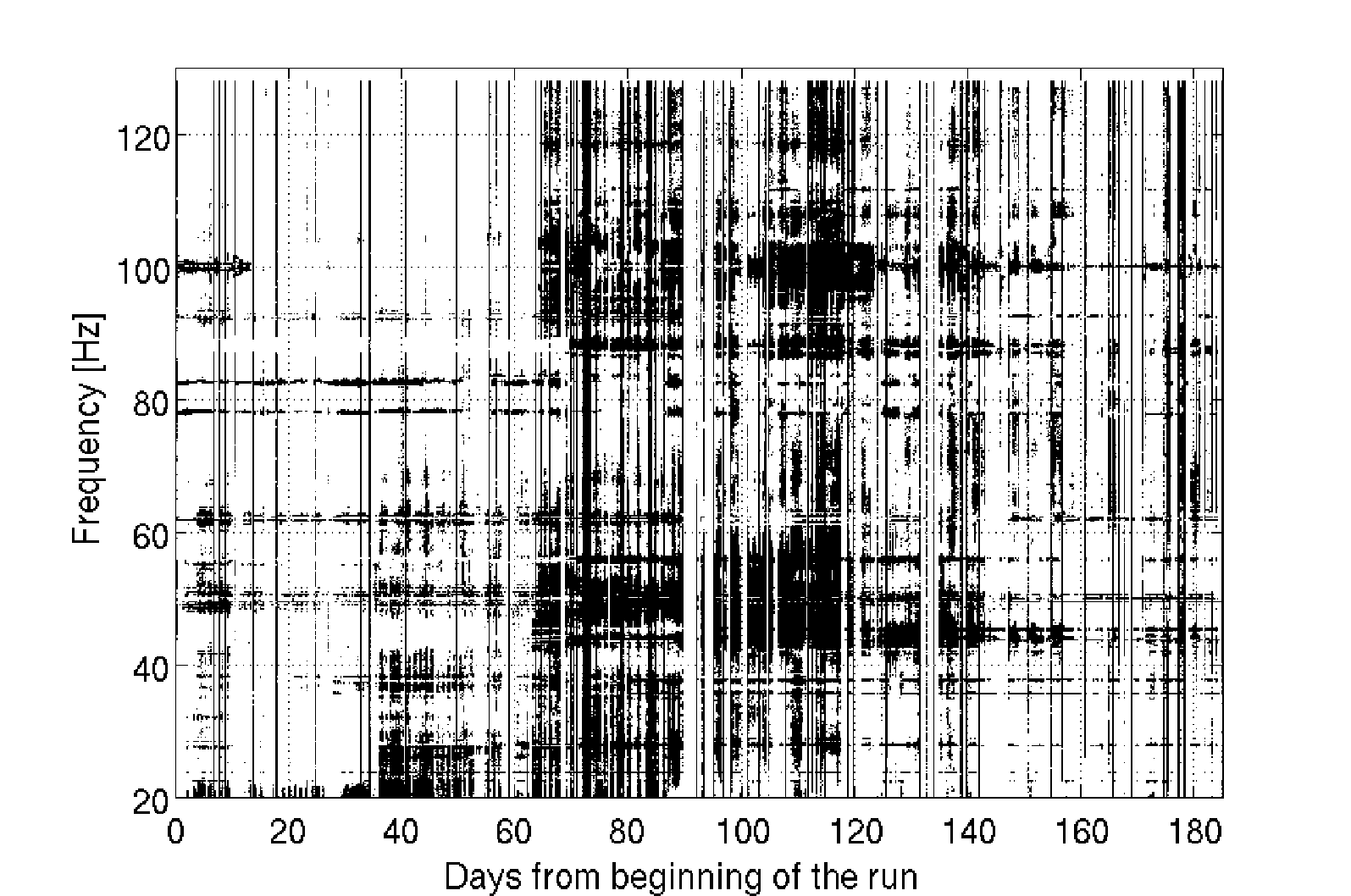} 
\includegraphics[width=8cm] {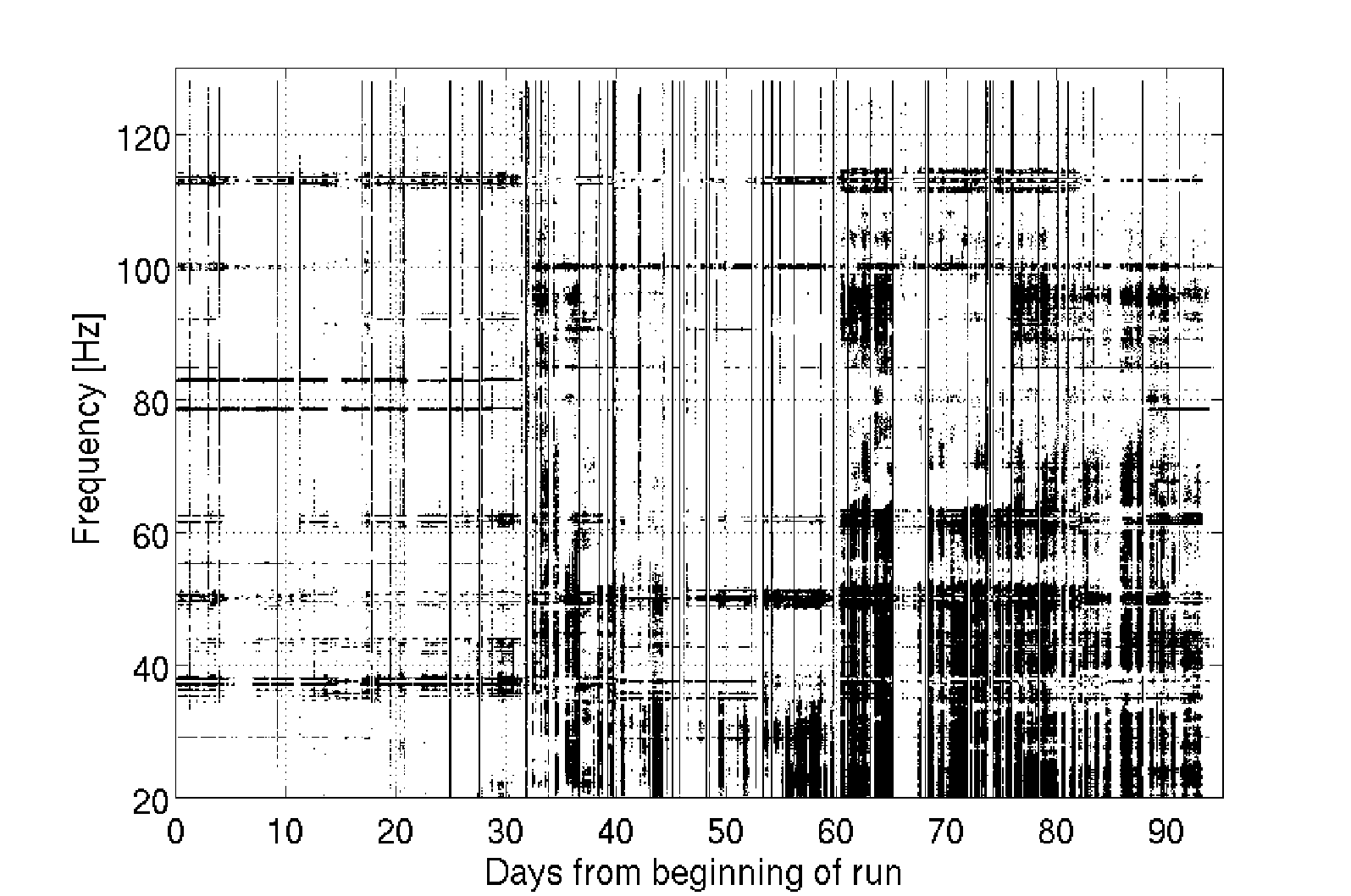}
\caption{Time-frequency plot of the peaks removed by the ``gross histogram'' cleaning procedure for VSR2 (left) and VSR4 (right).}
\label{fig:Grosshistveto_TF}
\end{figure*}

\begin{figure*}[!htbp]
\includegraphics[width=8cm]{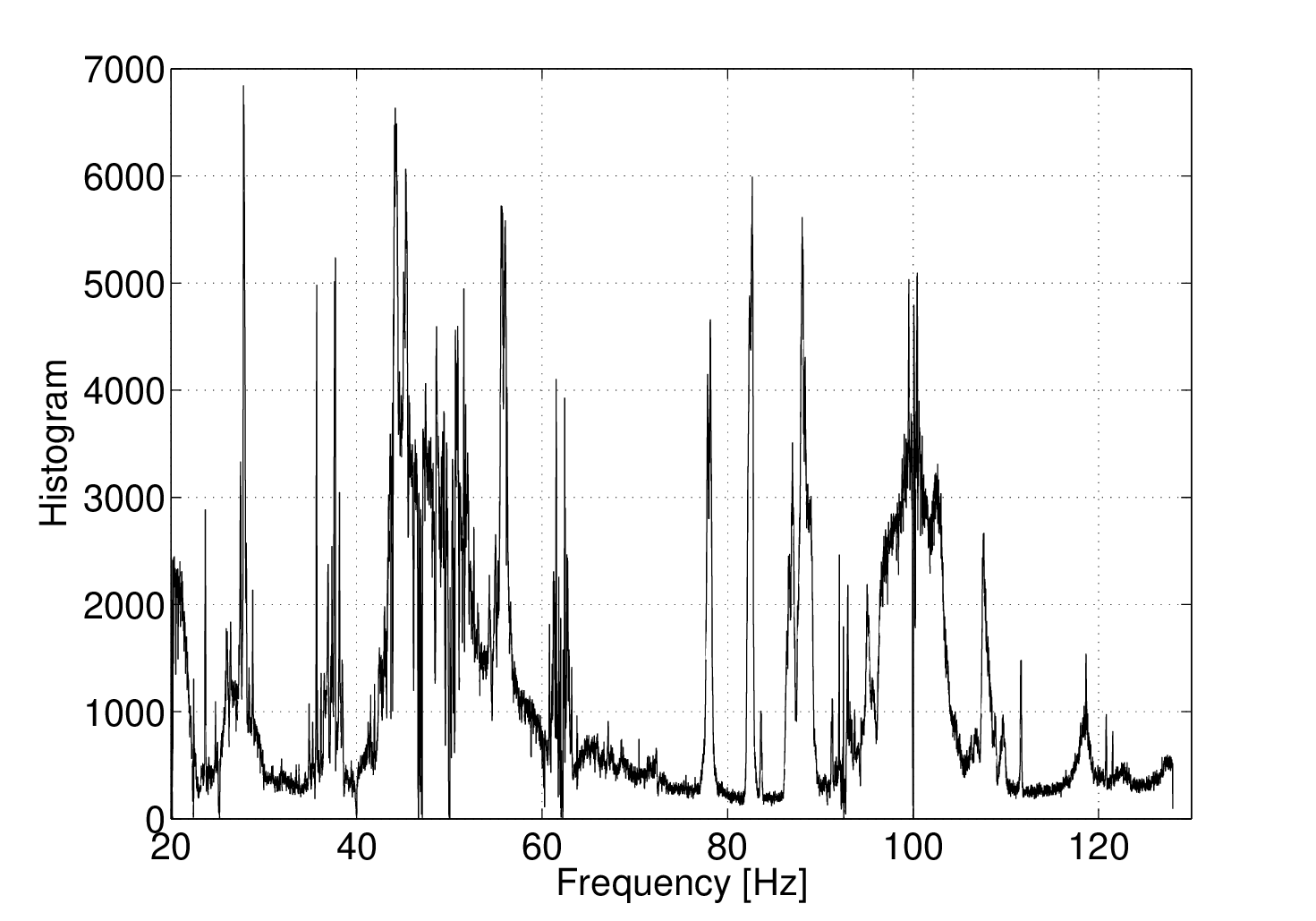} 
\includegraphics[width=8cm] {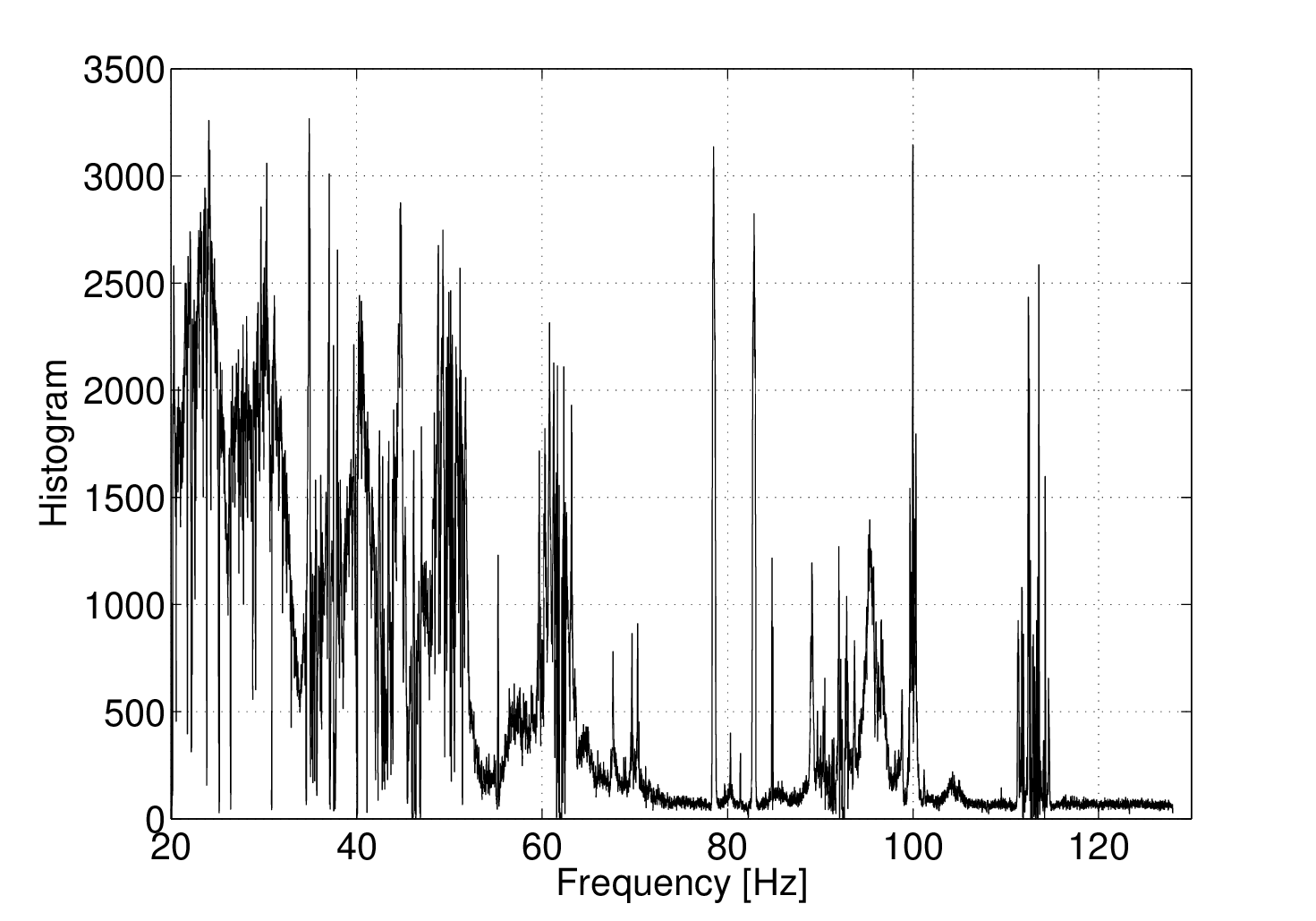}
\caption{Histogram of the peaks removed by the ``gross histogram'' veto for VSR2 (left) and VSR4 (right). The size of the bins in the histogram is 10 mHz.}
\label{fig:Grosshistveto_hist}
\end{figure*}

Spectral wandering lines and spectral lines of constant frequency are removed from the peakmaps \cite{ref:freqhoughmethod}.
Removal of spectral wandering lines (composed by peaks occurring at varying frequencies) is based on the construction of a histogram of low resolution (both in time and in frequency) peakmap entries, which we call the ``gross histogram''.
Based on a study on VSR2 and VSR4 data, we have chosen a time resolution of 12 hours and a frequency resolution of 0.01 Hz. In this way any true CW signal of plausible strength would be completely 
confined within one bin, but would not significantly contribute to the histogram, avoiding veto. As an example, Fig. \ref{fig:Grosshistveto_TF} shows the time-frequency plot of the 
peaks removed by the ``gross histogram'' cleaning procedure on VSR2 and VSR4 data and Fig. \ref{fig:Grosshistveto_hist} shows the histogram of the removed peaks, using a 10 mHz bin width, again both for 
VSR2 and VSR4 data. 

\begin{figure*}[!htbp]
\includegraphics[width=8cm]{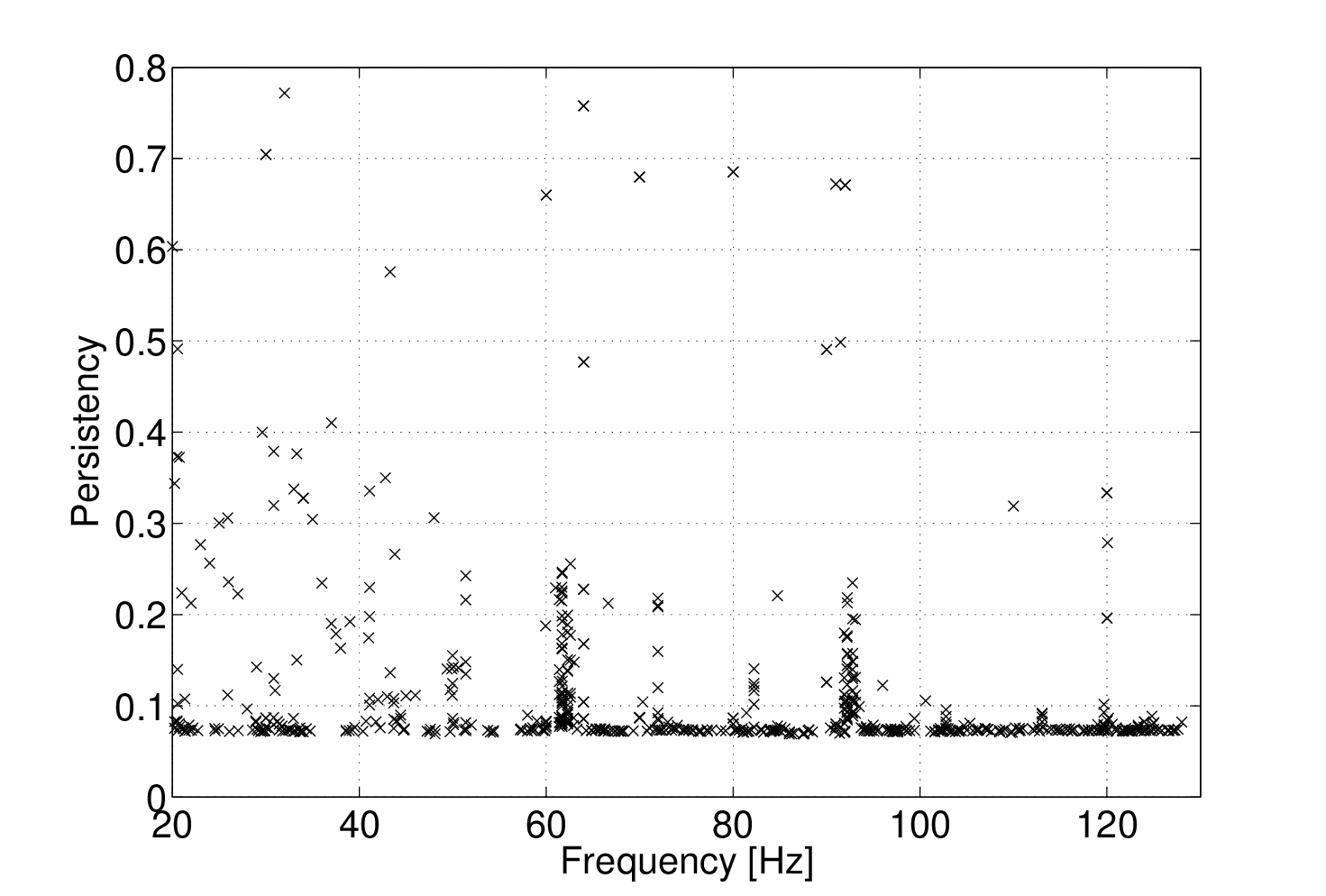} 
\includegraphics[width=8cm] {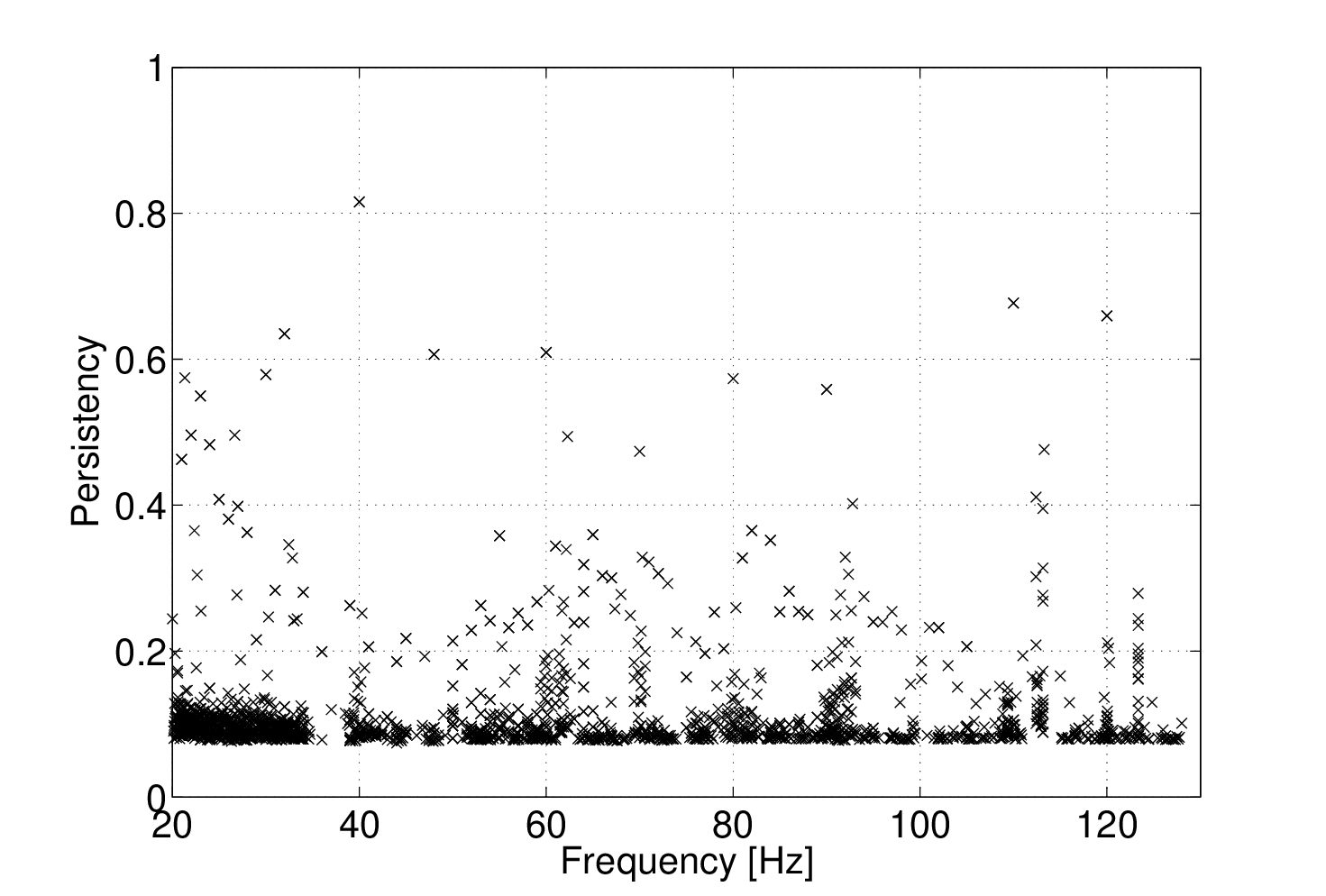}
\caption{Lines of constant frequency vetoed on the basis of the persistency, shown in the ordinate axis, for VSR2 (left) and VSR4 (right). We have removed 710 lines for VSR2 and 1947 lines for VSR4.)}
\label{fig:linesvetoed}
\end{figure*}

A second veto, aimed at removing lines of constant frequency, is based on the ``persistency'' analysis of the peakmaps, defined as the ratio between the number of FFTs in which a given line was present and the total number of analyzed FFTs. The vetoing procedure consisted of
histogramming the frequency bins and setting a reasonable threshold to select lines to be removed. To evaluate the veto threshold, we have used a ``robust'' statistic, described in Appendix D of \cite{ref:freqhoughmethod}, which is based on the median rather than the mean, which is much less affected by tails in the distribution. The resulting threshold on the persistency is of the order of 10\%. Figure \ref{fig:linesvetoed} shows the lines of constant frequency vetoed on the basis of the persistency, given on the ordinate axis. In this way we have vetoed 710 lines for VSR2 and 1947 lines for VSR4, which we know to be more disturbed than VSR2.
The ``gross'' histogram veto reduced the total number of peaks by 11.3\%\ and 13.1\%\ for VSR2 and VSR4 respectively. The persistency veto brings the fraction of removed peaks to 11.5 \% and 13.5 \% for VSR2 and VSR4 respectively. We end up with a total number of peaks which is 191,771,835 for VSR2 and 93,896,752 for VSR4.\newline
The use of the adaptivity in the FrequencyHough transform acts to reduce the effect of some of the remaining peaks, those corresponding to an average level of the noise  higher than usual or to times when the  detector-source relative position is unfavourable.

For testing purposes, ten simulated CW signals, called {\it hardware injections}, have been injected during VSR2 and VSR4 runs, by properly acting on the mirrors control system (see Sec. \ref{sec:HI} for more details). While we have developed a method to effectively remove HIs from the peakmap, it has not been applied for the present analysis as their presence allowed testing the whole analysis chain. We will show results from HIs analysis in Sec. \ref{sec:HI}.

\section{Candidate selection}
\label{sec:cand}
As  outlined in Sec. \ref{sec:proc}, after the FrequencyHough transform has been computed for a given dataset, the first level of refined candidates is selected.
Their number is chosen as a compromise between a manageable size and an acceptable sensitivity loss, as explained in Sect. IX of \cite{ref:freqhoughmethod}. Moreover, candidates are selected in a way to reduce as much as possible the effect of disturbances.  
Let us describe the selection procedure. The full frequency range is split into 1 Hz sub-bands, each of which is analyzed separately and independently.
Following the reasoning in  Sec. VIII of \cite{ref:freqhoughmethod}, we fixed the total number of candidates to be selected in each run to $N_{\mathrm{cand}}=10^8$. 
We distribute them in frequency by fixing their number in each 1 Hz band, $N_{\mathrm{cand},\,i}$, where $i=20,\ldots,127$, in such a way to have the same expected number of coincidences in all the bands. 
Moreover, for each given frequency band we have decided to have the same expected number of coincidences for each sky cell. Then, the number of candidates to be selected for each 1 Hz band and for each sky cell is given by $N^{\mathrm{sky}}_{\mathrm{cand},\,i}= N_{\mathrm{cand},\,i}/N_{\mathrm{sky},\,i}$, where $N_{\mathrm{sky},\,i}$ is the number of sky patches in the i$th$ frequency band.
Since $N_{\mathrm{cand},\,i}$ linearly increases with the frequency (see Eq. 48 in \cite{ref:freqhoughmethod}) and, for a given frequency band, the number of points in the sky increases with the square of the band maximum frequency, we have that the number of candidates selected per sky patch decreases with frequency. 
 
\begin{figure*}[!htbp]
\includegraphics[width=8cm]{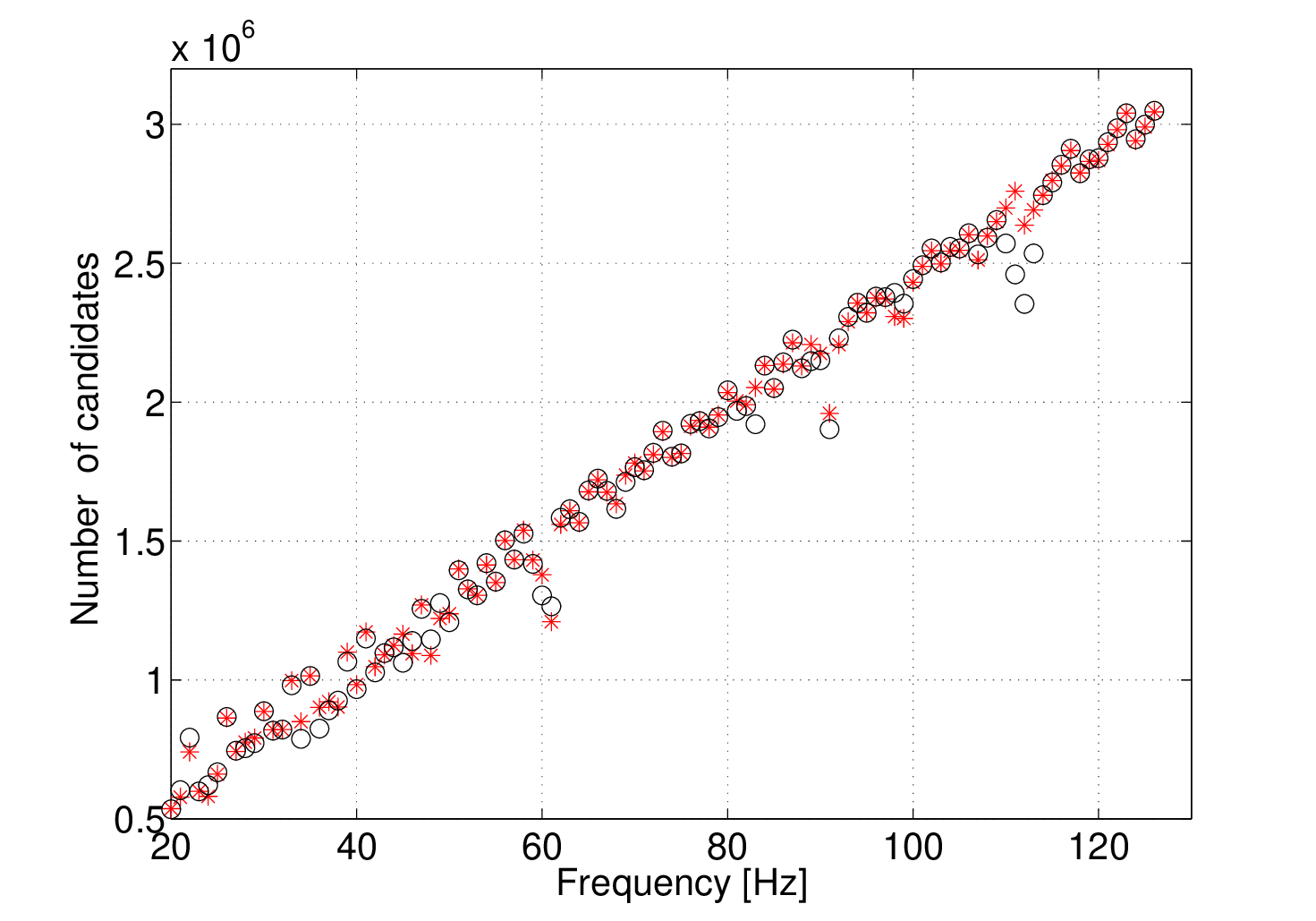}
\includegraphics[width=8cm]{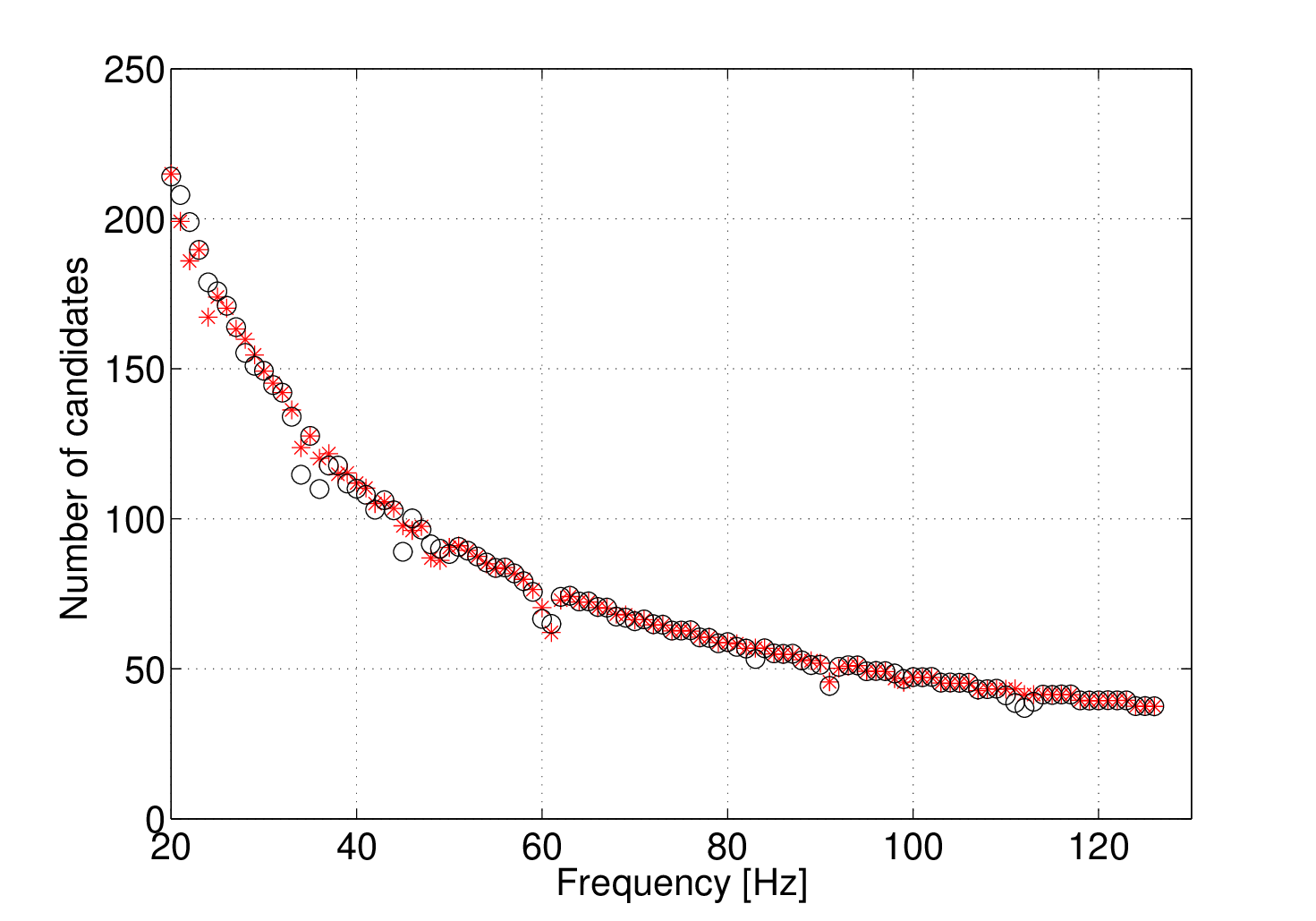}
\caption{Number of candidates selected in every 1 Hz band (left) and in every 1 Hz band and in every sky patch (right). Asterisks (red in the online version) corresponds to VSR2, black circles to VSR4.}  
\label{fig:ncandperband}
\end{figure*}

We now focus on some practical aspects of the procedure. As previously explained, we have fixed the size of the frequency bands to 1 Hz, the total number of 
candidates to be selected in each of them ($N_{\mathrm{cand},\,i}$) and
the number of candidates in each cell of the sky ($N^{\mathrm{sky}}_{\mathrm{cand},\,i}$). We have now the problem of being affected by the presence of disturbances of unknown origin. 
They could still pollute sub-bands in the $i$th band, even after performing all the cleaning steps described in Sec.\ref{sec:clean}. 
We have designed a procedure for candidate selection for this purpose \cite{ref:freqhoughmethod}. 
For each sky cell, we divide the $i$th band  
into $n_{\mathrm{sb}}=N^{\mathrm{sky}}_{\mathrm{cand},\,i}$ sub-bands, and select the most significant candidate in each of them, i.e. the bin in the Hough map with the highest amplitude. In this way a uniform selection 
of candidates is done in every band and the contribution of possible large disturbances is strongly reduced.
A further step consists of selecting the ``second order'' candidates. Once 
the most significant candidate in each sub-band has been selected, an empirically established exclusion region of $\pm 4$ frequency bins around it is imposed. This means we do not select any further candidate which is within that range from the loudest candidate in the sub-band. In this way we reduce the probability to select more candidates which could be due to the same disturbance. We now look for the second loudest candidate in that sub-band and select it only if it is well separated in frequency
from the first one, by at least $\pm 8$ frequency bins. In this way, we expect 
to select two candidates per sub-band in most cases and to have one candidate only when the 
loudest candidate is due to a big disturbance, or a particularly strong hardware injection.
Results are shown in Fig. \ref{fig:ncandperband} (left), which gives the number of candidates selected in every 1 Hz band and in Fig. \ref{fig:ncandperband} (right), 
which gives the number of candidates selected in every 1 Hz band and for each sky patch.
In the end we have 194,457,048 candidates for VSR2 and 193,855,645 for VSR4, which means that we have selected also the ``second order'' candidates in $\sim 96 \%$ of the cases.

\section{Candidate clustering and coincidence}
\label{sec:clusters}
We expect that nearly all the candidates selected in the analysis of a dataset will be false, arising from noise. In order to reduce the false alarm probability, coincidences among the two 
sets of candidates, found in VSR2 and VSR4 analysis, have been 
required. Indeed, given the persistent nature of CWs, a candidate in a dataset due to a real signal will (approximately) have the same parameters in another dataset, even if this 
covers a different time span. If a candidate is found to be coincident among the two datasets, it will be further analyzed as described in Sec. \ref{sec:followup}. 

In fact, due to
computational reasons, candidates from each dataset are {\it clustered} before making coincidences. 
A cluster is a collection of candidates such that the distance $d$ in the parameter space among any two is smaller than a threshold $d_\mathrm{clust}$. We have used an empirically chosen value $d_\mathrm{clust}=2$. 
Each candidate is defined by a set of 4 parameters: position in ecliptic coordinates $(\lambda,\beta)$, frequency $f$ and spin-down $\dot{f}$. Therefore, for two given candidates with $\vec{c}_1=(\lambda_1,~\beta_1,~f_1,~\dot{f}_1)$ and $\vec{c}_2=(\lambda_2,~\beta_2,~f_2,~\dot{f}_2)$ respectively, we define their distance as
\begin{equation}
d=\|\vec{c}_1-\vec{c}_2\|=\sqrt{k^2_{\lambda}+k^2_{\beta}+k^2_f+k^2_{\dot f}}.
\label{norm}
\end{equation}
Here $k_{\lambda}=|\lambda_2-\lambda_1|/\delta \lambda$ is the difference in number of bins between the ecliptic longitudes of the two candidates, 
and $\delta \lambda=(\delta \lambda_1+\delta \lambda_2)/2$ is the mean value of the width of the coarse bins in the ecliptic longitude for the two candidates (which can vary, as the resolution in longitude depends on the longitude itself), and similarly for the other terms. 
We find 94,153,784 clusters for VSR2 and 38,953,404 for VSR4.

Once candidates of both datasets have been clustered, candidate frequencies are referred to the same epoch, which is the beginning of VSR4 run. This means that the frequency of VSR2 candidate are shifted according to the corresponding spin-down value. Then coincidence requirements among clusters are imposed as follows. For each cluster of the first set, a check is performed on clusters in the second set.
First, a cluster of the second set is discarded if its parameters are not compatible with the trial cluster in the first set. As an example,  if the maximum frequency found in a cluster is smaller than the minimum frequency found in the other one, they cannot be coincident. 

For each candidate of the first cluster of a potentially coincident pair, the distance from the candidates of the second cluster 
is computed. If a distance smaller than $d_\mathrm{coin}=2$ is found, then the two clusters are considered coincident: the distance between all the pairs of candidates of the two clusters is 
computed and the pair with the smallest distance constitutes a pair of coincident candidates. This means that each pair of coincident clusters produces just one pair of coincident candidates.  
The choice $d_\mathrm{coin}=2$, based on a study of software-injected signals, allows us to reduce the false alarm probability and is robust enough against the fact that true signals can be found with
slightly different parameters in the two datasets.  
The total number of coincidences we found is 3,608,192.

At this point, coincident candidates are divided into bands of 0.1 Hz and subject to a {\it ranking} procedure in order to select the most significant ones.
Let $N$ be the number of coincident candidates in a given 0.1 Hz band. Let us order them in descending order of Hough map amplitude (i.e. from the highest to the smallest), separately for VSR2 and VSR4.
We assign a rank to each of them, which is $1/N$ to the highest and 1 to the smallest. We have then two sets of ranks, one for VSR2 candidates and one for VSR4 candidates.
Now we make the product of the ranks of coincident candidates.
The smallest possible value is $1/N^2$,
if a pair of coincident candidates has the highest Hough amplitude in both VSR2 and VSR4, while the largest possible value is 1, if a pair of coincident candidates has the smallest amplitude in both VSR2 and VSR4.
For each 0.1 Hz band we select the candidates having the smallest rank product. 
We chose 0.1 Hz wide bands as a good compromise between having a statistic large enough in every band and reducing the effect of (strong) noise outliers in wider bands, as will be clear in the following.
With this choice we have 1080 coincident candidates over the band 20-128 Hz. 

\begin{figure*}[!htbp]
\includegraphics[width=12cm]{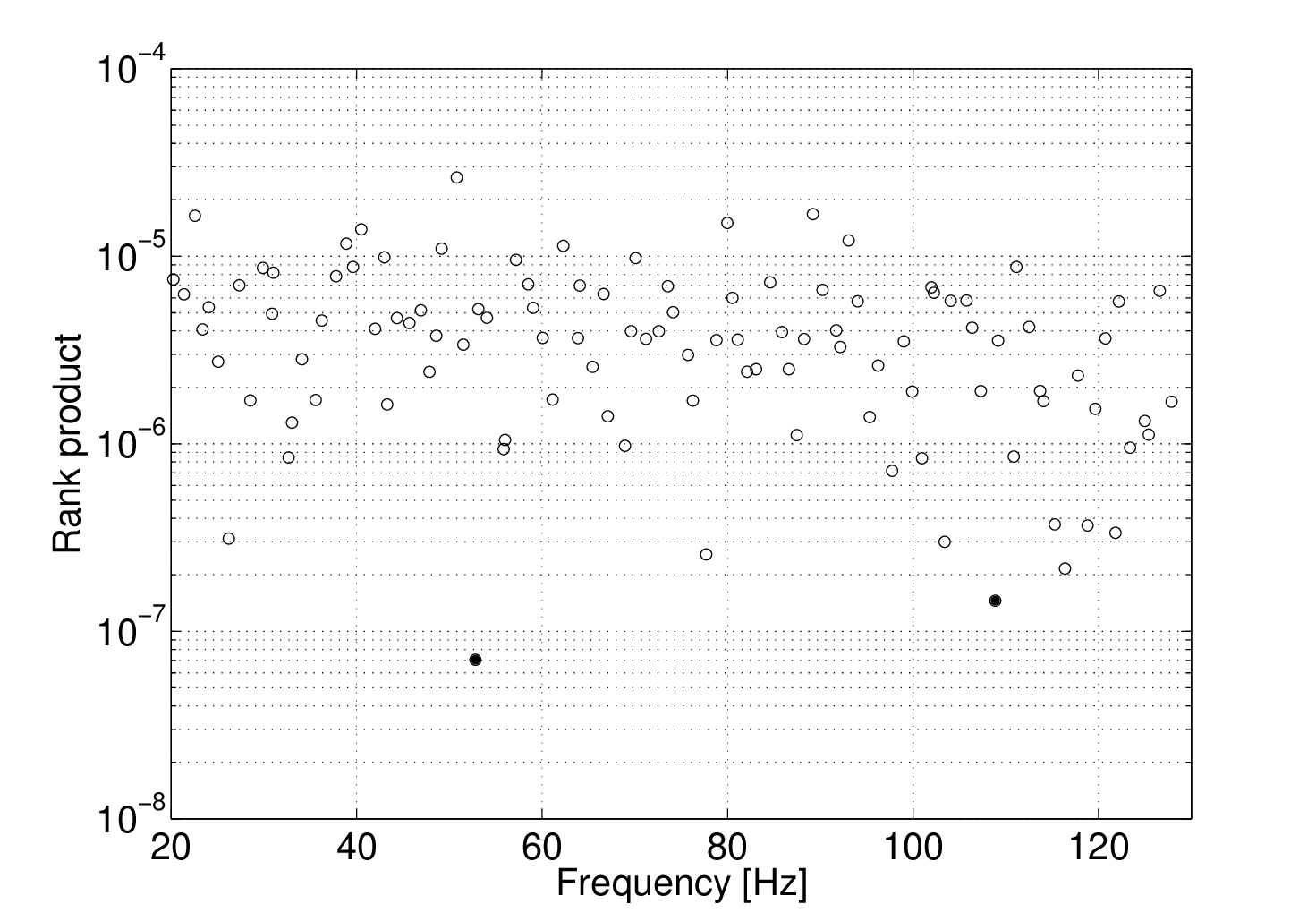}
\caption{Rank product for the final 108 candidates, which will be subject to the follow-up procedure. The two candidates with smallest rank product, indicated by filled circles, correspond to two hardware-injected signals.}
\label{fig:rank_product} 
\end{figure*}

At this point,
among the ten candidates in ten consecutive 0.1 Hz bands we chose the most significant one, i.e. that having the smallest rank product, ending up with 108 candidates, one per 1 Hz band, which will be subject to a {\it follow-up} procedure, as described in the 
next section. Note that due to the cleaning steps we apply, especially those on the peakmaps, more disturbed 0.1 Hz bands tend to have a smaller number of candidates and then a smaller number of coincidences $N$. This reduces the chance that the most significant candidate in a 1 Hz band comes from a particularly noisy 0.1 Hz interval. The number of candidates selected at this stage depends on the amount of available computing resources and is constrained by the amount of time the analysis will take. Figure \ref{fig:rank_product} shows the rank product for the final 108 candidates as a function of their frequency. The two smallest values corresponds to the hardware-injected signals at 52.8 Hz and 108.3 Hz. 
In Tab. \ref{tab:SELb} the main parameters of the 108 selected candidates are given. They have been ordered as a function of the value of the ranking parameter, starting from the most 
significant one.

\section{Candidate follow-up}
\label{sec:followup}
The follow-up procedure consists of several steps applied to each of the 108 selected candidates. The basic idea is that of repeating the previous incoherent analysis with an improved sensitivity, which can be obtained by increasing the time baseline of the FFTs. In order to do this a coherent step, using the candidate parameters, is done separately for the two runs at first. 
This is based on the same analysis pipeline used for targeted 
searches \cite{ref:vela_vsr2}, \cite{ref:vsr4_kp}, \cite{ref:nr_method}, \cite{ref:nr_obs} and the output is a down-sampled (at 1 Hz) time series where the Doppler effect, the spin-down and the Einstein delay for a source, 
having the same parameters as the candidate, have been corrected. A final cleaning stage is also applied, by removing the non-Gaussian tail of the data distribution. 

From these data, a new set of longer FFTs is computed, from chunks of data of duration 81,920 seconds, which is ten times longer than in the initial step of the analysis. This procedure preserves true signals with parameters slightly different from those of the candidate, but within the uncertainty window because of the partial Doppler and spin-down correction just applied. If we assume that the true signal has a frequency different from that of the candidate by a coarse bin $1/T_{\mathrm{FFT}}$, and a sky position also shifted by a coarse sky bin, the maximum allowed FFT duration can be numerically evaluated using Eq.~36 in \cite{ref:freqhoughmethod}, and is of about 290,000 seconds, significantly larger than our choice. 

From the set of FFTs a new peakmap is also computed, selecting peaks above a threshold of $\sqrt{5.5}\simeq 2.34$ on the square root of the equalized spectra, significantly higher than 
that (1.58) initially used (see Sec. \ref{sec:proc}). This is justified by the fact that a signal present in the data, and strong enough to produce a candidate, would contribute similarly when lengthening the FFT (by a factor of 10) while the noise in a frequency bin
decreases by a factor of 10 (in energy). In fact, 
a threshold of 2.34 is a very conservative choice which, nevertheless, reduces the expected number of noise peaks by a factor of 20 with respect to the initial threshold.       

At this point the peakmaps computed for the two runs are combined, so that their peaks cover a total observation time of 788.67 days, from the beginning of VSR2 to the end of VSR4 
run. Before applying the incoherent 
step, a grid is built around the current candidate. The frequency grid is built with an over-resolution factor of 10, as initially done, so that $\delta f=1/81920\times 10=1.22\times 
10^{-6}$ Hz, and covers 0.1 Hz around the candidate 
frequency. The grid on the spin-down is computed with a step 10 times smaller than the coarse step of the shorter run (VSR4), and covering $\pm 1$ coarse step, so that 21 spin-down bins are considered. 
Finally, a grid of 41 by 41 sky points (1681 in total) is built around the candidate position, covering $\pm 0.75$ times the dimension of the coarse patch (whose 
extension actually depends on the position in the 
sky and on the candidate frequency). For every sky point the peakmap is Doppler corrected by shifting each peak by an amount corresponding to the Doppler shift, and a FrequencyHough transform is 
computed. This means that for each candidate 41$\times$41=1681 FrequencyHough transforms are computed. The loudest candidate, among the full set of FrequencyHough maps, is then selected. 

The starting peakmap is now corrected using the parameters of the loudest candidate and projected on the frequency axis. On the resulting histogram we search for significant peaks. This is done by taking the maximum of the histogram over a search band covering a range of $\pm 2$ coarse bins around the candidate frequency. The rest of the 0.1 Hz band is used to estimate the background noise: it is similarly divided in sub-intervals covering 4 coarse bins and the loudest candidate is taken in each of them. In total we have 408 sub-intervals. In the frequentist approach the significance of a candidate is measured by the p-value, that is the probability that a candidate as significant as that, or 
more, is due to noise. The smaller is the computed p-value and less consistent is the candidate with noise. We estimate a p-value associated to the candidate by computing the fraction of sub-bands in which the loudest peak is larger than that in the search band. The smaller is the p-value, the higher is the candidate significance. If the p-value is smaller than 1$\%$ the candidate can be considered as ``interesting'', namely it  is not fully compatible with noise, and will be subject to a deeper scrutiny, looking at the consistency of the candidate between the two runs and, possibly, extending the analysis to other data sets or applying a further step of follow-up with a longer coherence time. 

\begin{figure*}[!htbp]
\includegraphics[width=12cm]{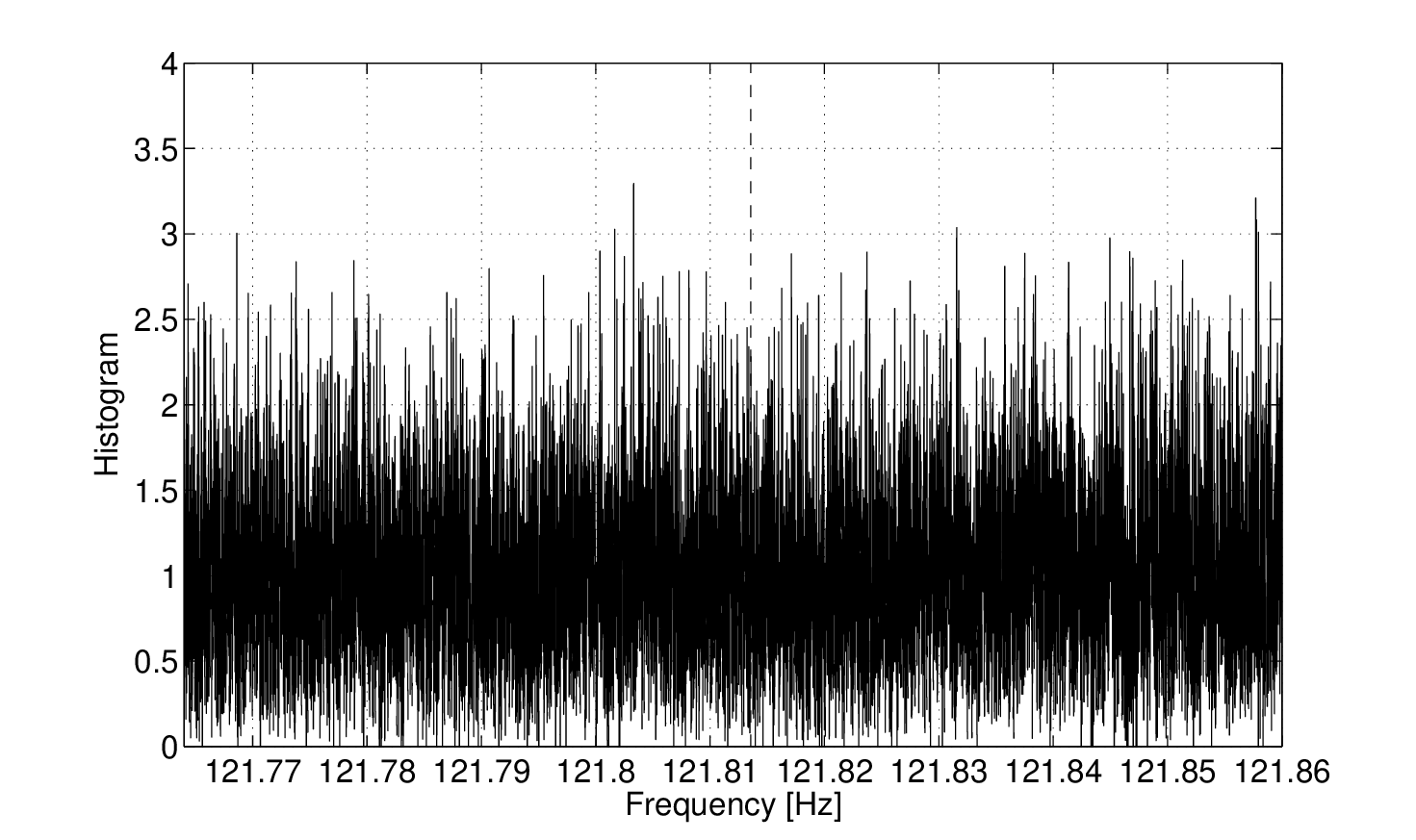}
\caption{Peakmap histogram around the candidate at frequency 121.81 Hz. The vertical line identifies the candidate frequency.}  
\label{fig:cand121hist}
\end{figure*}

As an example, Fig. \ref{fig:cand121hist} shows the peakmap histogram for the candidate at frequency 121.81 Hz, in which no significant peak is found around the candidate frequency (identified by the vertical dashed line). The corresponding p-value is $\sim$0.44. 

In contrast, Fig. \ref{fig:hi_hist} shows the peakmap histograms for the candidates at 52.80 Hz (left), corresponding to hardware injection {\it pulsar\_5}, and
the candidate at 108.85 Hz (right), corresponding to hardware injection {\it pulsar\_3}. In each case a highly significant peak is visible at the frequency of the candidate (in fact it is the largest peak in the whole 0.1 Hz band around it), as detailed in Sec. \ref{sec:HI}. 

\begin{figure*}[!htbp]
\includegraphics[width=12cm]{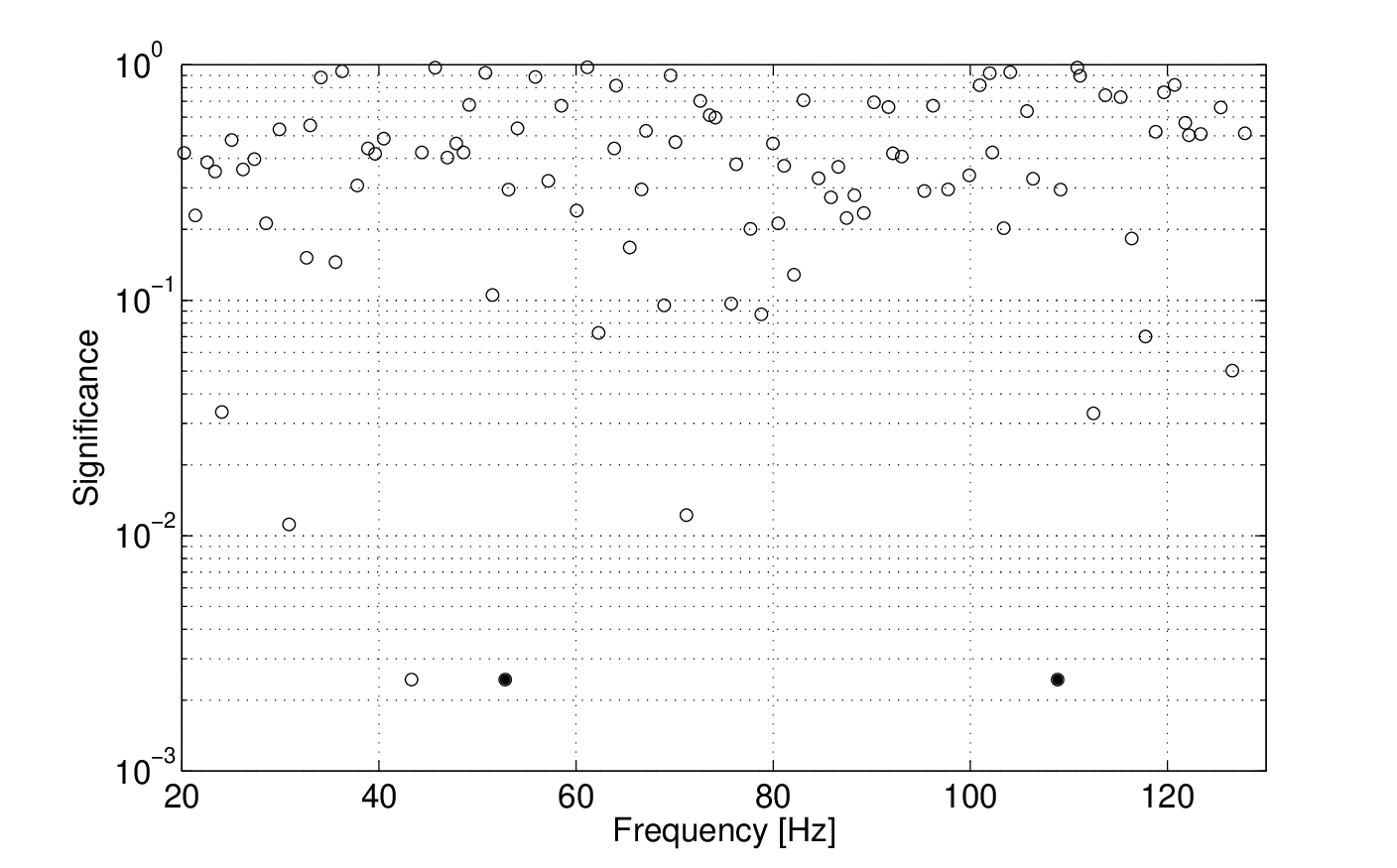}
\caption{Final p-values for the 108 candidates as a function of candidate frequency after performing the follow-up step. The minimum value is 1/408=0.00245. The two filled circles identify hardware-injected signals.}  
\label{fig:cand_significance}
\end{figure*}

The p-value for the 108 candidates is given in the last column of Tab. \ref{tab:SELb}. In Fig. \ref{fig:cand_significance} the p-value for the 108 candidates is shown as a function of candidate frequency. 
Both the hardware injections have the lowest possible p-value (equal to 1/408=2.45$\times 10^{-3}$), that is the highest possible significance, providing a good check that the full analysis 
pipeline works correctly. 

Moreover, 
there are three outliers, which are candidates not associated to injected signals, having a p-value very close or below the 1$\%$ threshold we have established. These three candidates have undergone further analysis in order to increase their significance or to discard them as possible GW signals. In fact, as described in Appendix \ref{sec:outl}, all of them appear to be incompatible with GW signals.
    
\section{Hardware injection recovery}
\label{sec:HI}
\begin{figure*}[!htbp]
\includegraphics[width=8cm]{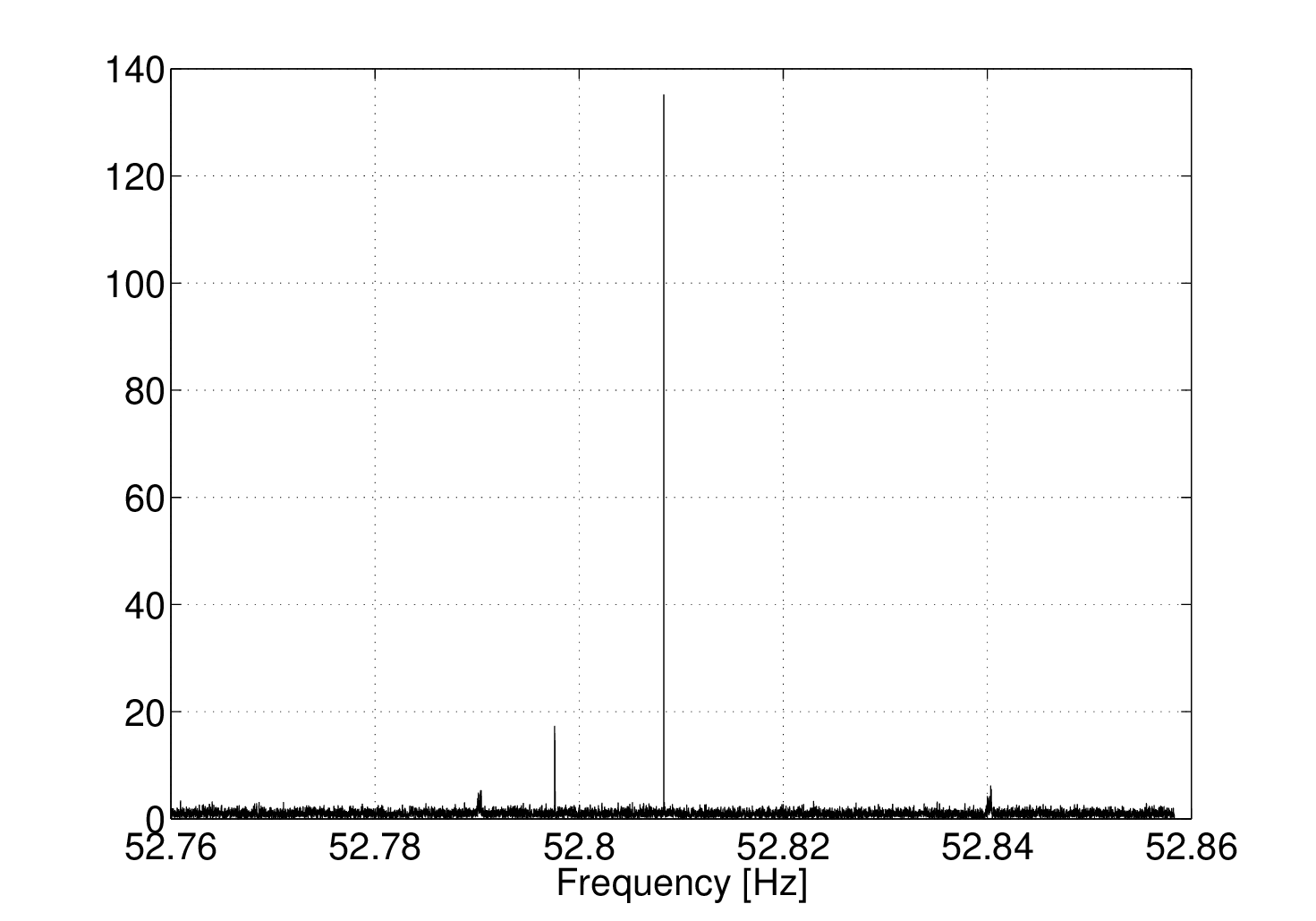}
\includegraphics[width=8cm]{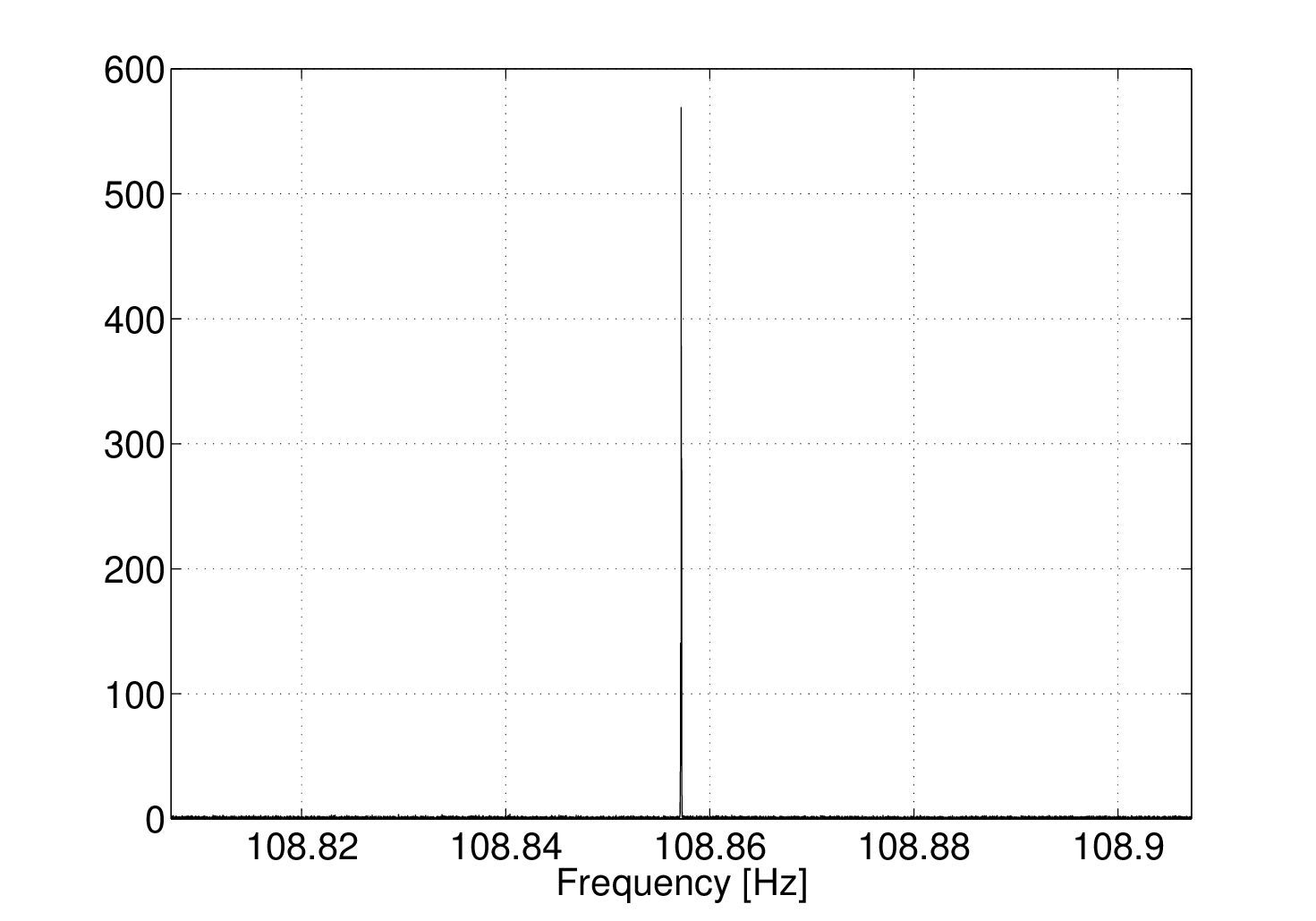}
\caption{Peakmap histogram around the hardware injections {\it pulsar\_5} (left) and {\it pulsar\_3} (right). In both cases the highest peak corresponds to the signal frequency, with high accuracy.}  
\label{fig:hi_hist}
\end{figure*}

Ten simulated CW signals have been continuously injected in the Virgo detector during VSR2 and VSR4 runs, by 
actuating on the mirror control system with the aim of 
testing analysis pipelines. Two of these {\it hardware} injections have frequency below 128 Hz and have 
been considered in this work:
{\it pulsar\_5}, with a frequency of about 52.8 Hz, and {\it pulsar\_3}, at about 108.85 Hz.
In Tab. \ref{tab:HI} parameters relevant for the current analysis are given.
\begin{table}[!htbp]
\begin{center}
\begin{tabular}{|c|c|c|c|c|c|}
\hline
 HI   & $f_{\rm gw}$  & $\dot f $ & $\lambda$  & $\beta$ & $H_0$ \\
       &  [Hz]         & [Hz/s]     & [deg]           & [deg]      & $\times 10^{-24}$ \\
\hline
{\it pulsar\_5}  & 52.80832  & $-4.03 \times 10^{-18}$ &276.8964 & -61.1909 & 3.703 \\
\hline
{\it pulsar\_3}  & 108.85716  & $-1.46 \times 10^{-17}$ &193.3162 & -30.9956 & 8.296 \\
\hline
\end{tabular}
\caption{Main parameters of the hardware injections in the band of the present analysis. The reference epoch for the frequency is 
MJD 52944.}
\label{tab:HI}
\end{center}
\end{table}
These signals are strong enough to appear as the most significant candidates in the final list (Tab. \ref{tab:SELb}).
As such they have been also subject to the follow-up procedure. In Fig. \ref{fig:hi_hist} we show the final peakmap 
histograms for the candidate corresponding to {\it pulsar\_5} (left) and  {\it pulsar\_3} (right), where in both cases a very high peak is clearly visible in correspondence of the signal frequency. This is reflected in the high significance level shown in Tab. \ref{tab:SELb}. 
Tab. \ref{tab:fuhi} contains the estimated parameters for
the hardware injections, together with the associated error, which is the difference with respect to the injected values. For both signals the parameters are well recovered with an improved accuracy thanks to the followup. 
 \begin{table*}[!htbp]
\begin{center}
\begin{tabular}{|c|c|c|c|c|c|c|c|c|}
\hline
 HI   & $\hat{f}$ [Hz] & $\Delta f $ & $\hat{\dot f}$ [Hz/s] & $\Delta \dot{f}$ & $\hat{\lambda}$ [deg] & $\Delta \lambda$ & $\hat{\beta}$ [deg] & $\Delta \beta$ \\
\hline
{\it pulsar\_5}  & 52.80830 & $-$1.8 & 0.0 & $4\times 10^{-6}$ &276.93 & 0.12  & $-$61.19 & 0.007 \\
\hline
{\it pulsar\_3}  & 108.85714 & $-$1.2 & $1.18\times 10^{-12}$  & $1.05$ & 193.36 & $-$0.05 & -31.00 & 0.66 \\
\hline
\end{tabular}
\caption{Estimated parameters, and corresponding errors, for candidates corresponding to hardware injections present in the band bewteen 20 Hz and 128 Hz. The parameters $\hat{f}, \hat{\dot f}, \hat{\lambda}, \hat{\beta}$ are the estimated frequency, spin-down, ecliptic coordinates; $\Delta f, \Delta \dot{f}, \Delta \lambda, \Delta \beta$ are the differences with respect to the injected parameter values in units of the corresponding bins.}
\label{tab:fuhi}
\end{center}
\end{table*}
   
\section{Results}
\label{sec:ul}

As all the 106 non-injection candidates appear to be consistent with what we would expect from noise only, we proceed to compute an upper limit on the  signal amplitude in each
1-Hz band between 20 Hz and 128 Hz, with the exception of the band 52-53 Hz and 108-109 Hz, which correspond to the two hardware injections (in fact, see point 2. in this section, there are also several noisy sub-bands which have been excluded {\it a posteriori} from the computation of the upper limit). We follow a standard approach adopted for all-sky CW searches \cite{ref:ein@ho}, \cite{ref:alisky} and detailed in the following. Many (of the order of 200,000 in our case) simulated signals are injected into the data, and the signal strain amplitude is determined such that 90 $\%$ of the signals (or the desired confidence level) are detected, and are more significant than the original candidate found in the actual analysis of the same frequency band.

In other words, for each 1-Hz band we generate 100 simulated signals in the time domain, with unitary amplitude and having frequency and the other parameters drawn from a uniform distribution: the spin-down within the search range, position over the whole sky, polarization angle $\psi$ between  $-\frac{\pi}{2}$ and $\frac{\pi}{2}$, and inclination of the rotation axis with respect to the line of observation, $\cos{\iota}$, between $-$1 and 1. The time series containing the signals are then converted in a set of FFTs, with duration 8192 seconds, the same used for the analysis, and covering the band of 1 Hz. These ``fake signal FFTs'' are multiplied by an amplitude factor and summed to the original data FFTs. This is repeated using of the order of 10-15 different amplitudes, chosen in order to be around the expected 90$\%$ confidence level upper limit. 

\begin{figure*}[!htbp]
\includegraphics[width=8cm]{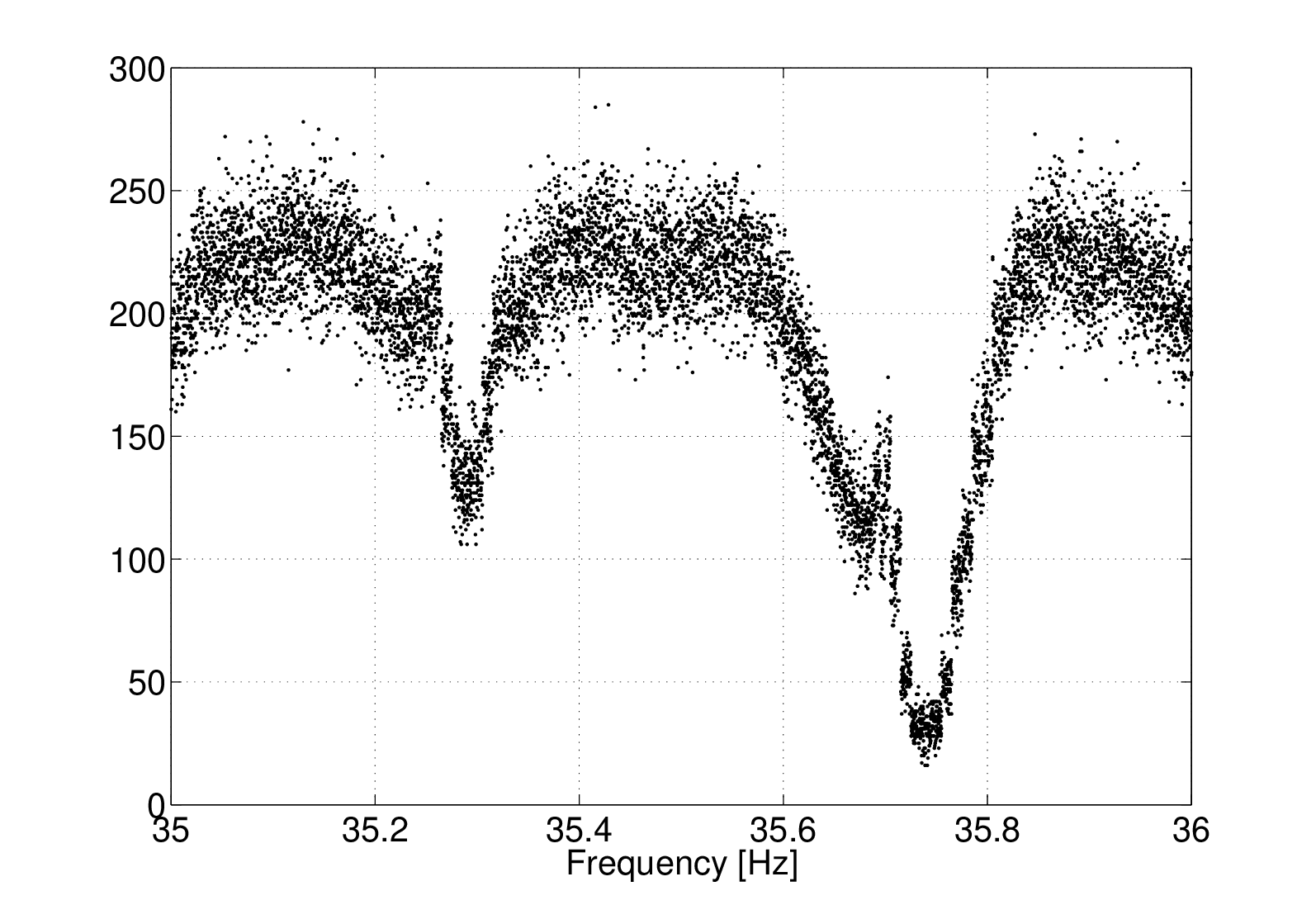}
\includegraphics[width=8cm]{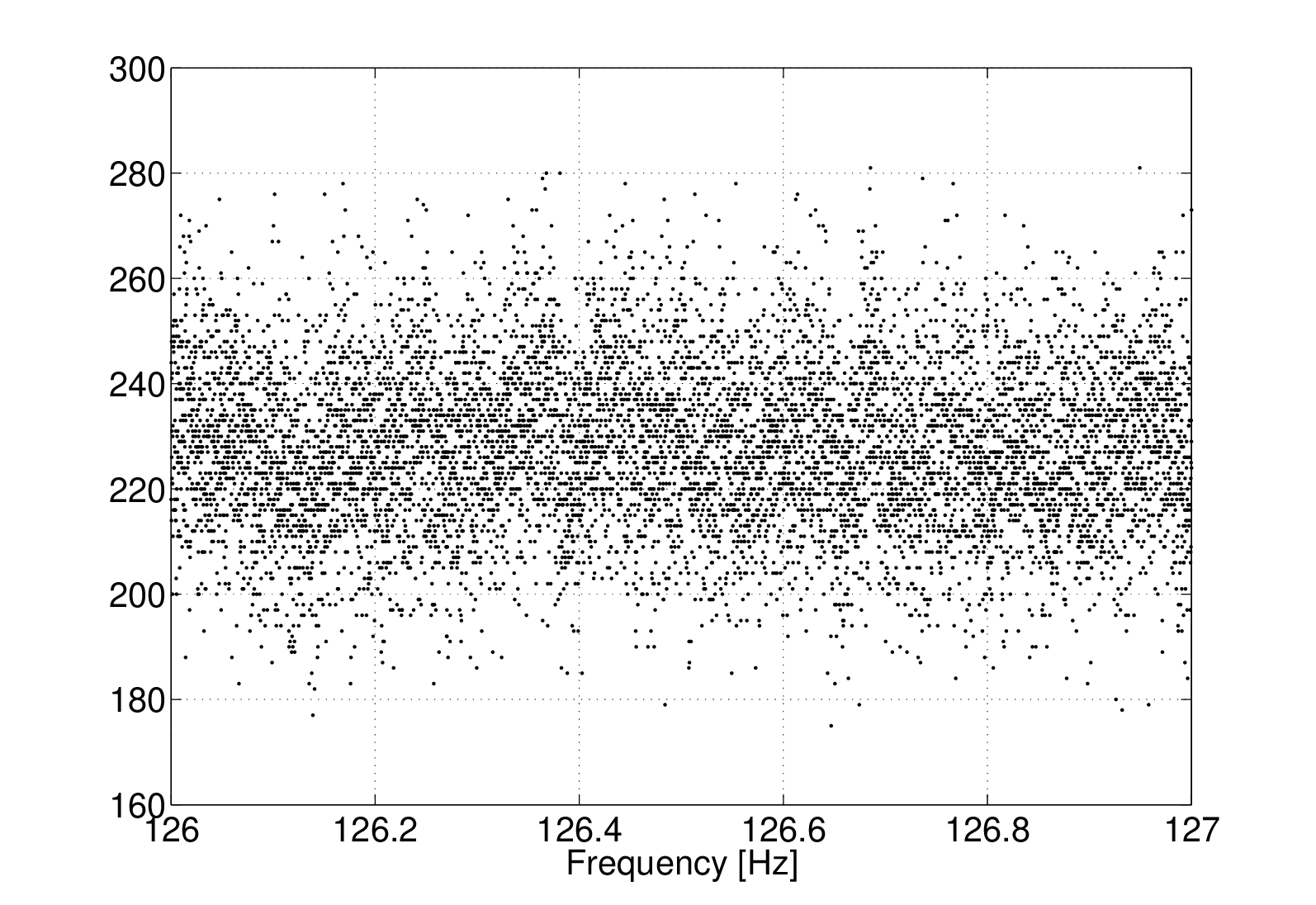}
\caption{Peakmap peak distribution for two bands in VSR2 data. The left plot refers to the band 35-36 Hz, with gaps clearly visible. The right plot refers to a much cleaner band, ranging from 126 Hz to 127 Hz.}  
\label{fig:holes}
\end{figure*}

For each set of 100 injections an analysis is done using the all-sky search code over a reduced parameter space around each injection, consisting of a frequency band of $\pm 0.2$ Hz around the injection frequency, the same spin-down range used in the production analysis, and 9 sky points around the coarse grid point nearest to the injection position. We have verified, by injecting software simulated signals into Virgo VSR2 and VSR4 data, that for signal amplitudes around the approximately expected upper limit values this volume is sufficiently large to contain all the candidates produced by a given signal, and at the same time is small enough to make the procedure reasonably fast. In fact, the frequency and position of the injections are chosen in such a way that any two signals are separated by at least 0.005 Hz and their sky search regions do not overlap. 
The output of this stage is, for each injected signal, a set of candidates from VSR2 data and another set of candidates from VSR4. In the production analysis, candidates of each run are clustered before making coincidences. For the computation of the upper limit the clustering cannot be applied because of the density of signals which would strongly affect the cluster composition introducing a mixing of candidates belonging to different injections. We recall that the  candidate clustering has been introduced in the production analysis with the main purpose to reduce the computational cost of the coincidence step. However, for the upper limit computation, a direct coincidence step among candidates is still affordable. Hence, for every candidate found in VSR2 we determine, if it exists, the nearest coincident candidate found in VSR4 with a distance smaller than 2. The coincident candidates are then ranked using the same procedure described in Sec.\ref{sec:clusters}, and the most significant one is selected. Then, for each injected signal amplitude we have at most 100 coincident candidates. They are compared to the candidate found in the actual analysis in the same 1-Hz band, in order to count the fraction ``louder'' than that. 
Two issues arise here and must be properly addressed. 
\begin{enumerate}
\item In principle, the comparison between the candidates found after injections and the actual analysis candidate should be done on the basis of their ranking. In fact this cannot be done because the numerical values of the ranking depend on the number of candidates present in the 1-Hz band, and this number is very different in the two cases, as the injection analysis is performed over a smaller portion of the parameter space. We have then decided to make the comparison using the Hough amplitudes of the candidates: a candidate found in the injection analysis is considered ``louder'' if the amplitudes of its parent candidates, found in VSR2 and VSR4 data, are both larger than the corresponding amplitudes of the parents of the actual analysis candidate. This will tend to slightly overestimate the upper limit. 
\item We have verified that in some bands the upper limit can be heavily affected by the presence of gaps in the starting peakmap, namely frequency bands in which the number of peaks is significantly smaller than in the rest of the band. As an example, in Fig. \ref{fig:holes} the peakmap peak distribution is shown for a ``bad'' frequency band between 35 and 36 Hz in VSR2 data, containing gaps, and for a good one, between 126 and 127 Hz. 
The gaps are the result of the various cleaning steps applied to the data, and the injections that, for a given amplitude, have frequency overlapping with a gap will not able, in most cases, to produce detectable candidates because the amplitude in the Hough map near the signal frequency is reduced by the smaller data contribution. As a consequence, the upper limit tends to be significantly worse with respect to the case in which the gaps are not relevant. To cope with this issue we opted to exclude from the computation of the upper limit the parts of a 1-Hz band that correspond to gaps. This procedure excludes the injections that happen to lie within the excluded region. In this way we obtain better results, which are valid only in a sub-interval of the 1-Hz band. In Appendix \ref{sec:excluded} we report all the intervals excluded from the upper limit computations. In total they cover about 8.6 Hz (out of the 108 Hz analyzed). 
\end{enumerate}
For each 1-Hz band, and for each injected signal amplitude, once we have computed the fraction of candidates louder than the actual analysis candidate, the pair of amplitudes that are nearest, respectively from above and from below, to the 90$\%$ confidence level are used as starting point of a new round, in which a new set of signals with amplitudes between those values are injected until the final confidence level of $90\%$ is reached. Overall, of the order of $2\times 10^5$ signals have been injected in both VSR2 and VSR4 data. The final accuracy of the upper limit depends on the number of injections actually used and on the quality of the data in the considered 1-Hz band. By repeating of the order of 10 times the upper limit computation in a few selected bands, we have verified that the upper limits have an accuracy better than $\sim$5$\%$ for ``bad'' bands and better than about $1\%$ for ``good'' bands, in any case smaller than the amplitude uncertainty due to the data calibration error (Sec. \ref{sec:ITF}).
   
\begin{figure*}[!htbp]
\includegraphics[width=12cm]{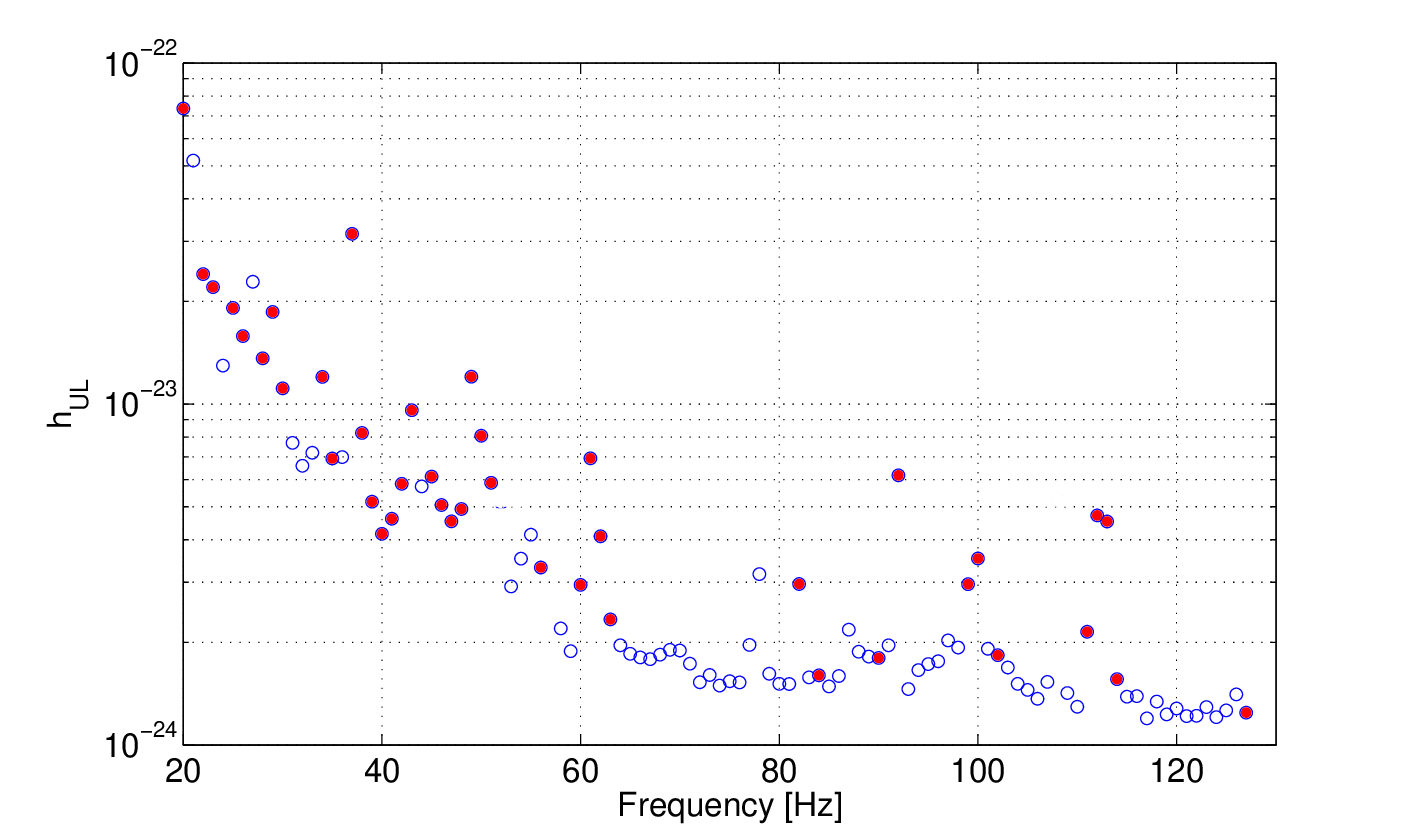} 
\caption{Joint VSR2-VSR4 upper limit on dimensionless strain as a function of the frequency. Open circles refer to upper limits values valid over the full corresponding 1-Hz band, while filled circles refer to upper limit values valid only in a portion of the corresponding 1-Hz band.}
\label{fig:ul_full}
\end{figure*}

\begin{figure*}[!htbp]
\includegraphics[width=12cm]{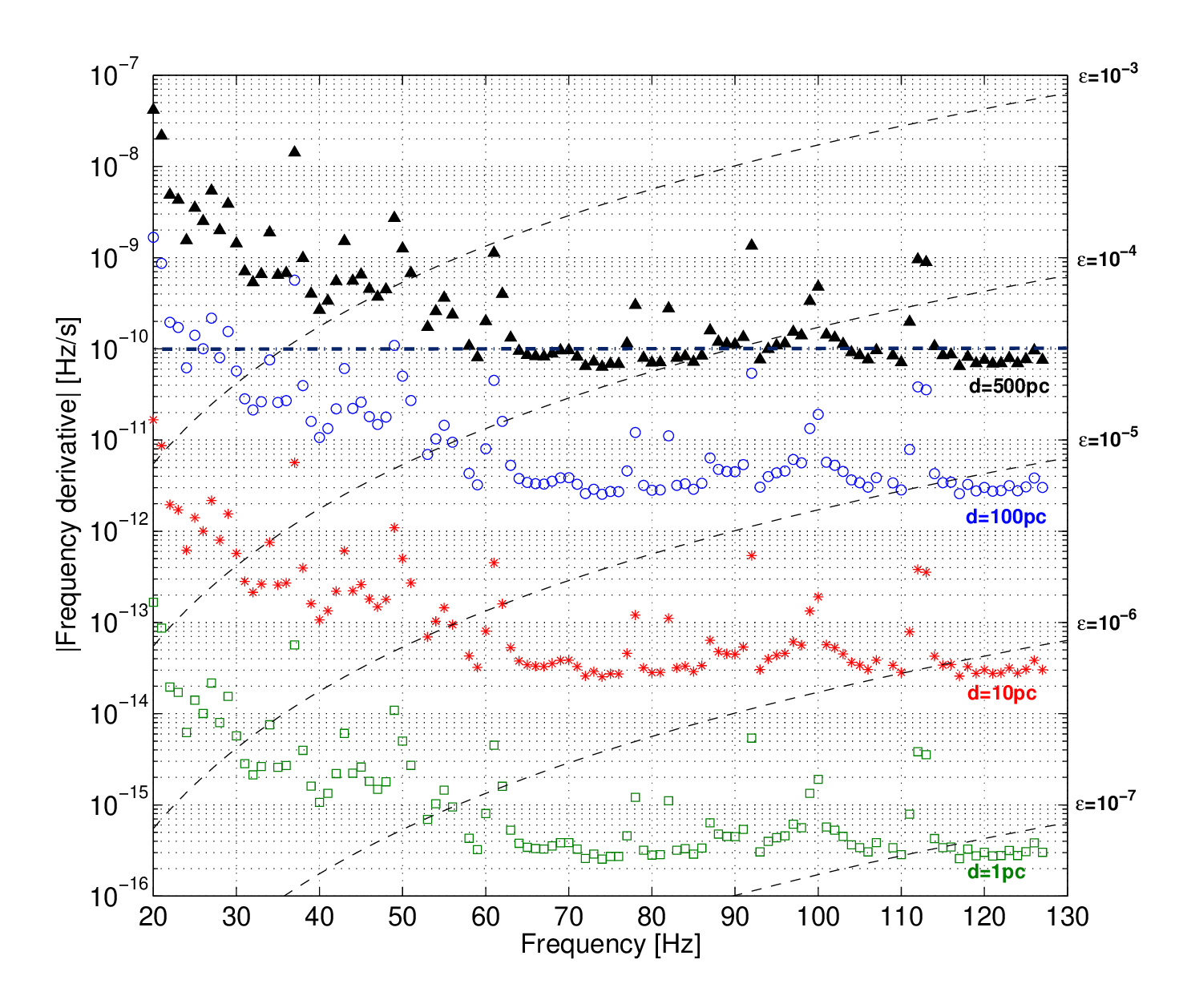} 
\caption{Astrophysical reach of the search. The sets of points gives the relation between the frequency derivative and the frequency of a signal emitted by a detectable source placed at various 
distances. The triangles correspond to a distance of 500 pc; the circles to a distance of 100 pc; the stars to a distance of 10 pc and the squares to a distance of 1 pc. The dashed lines 
represents lines of constant ellipticity. The horizontal dot-dashed line indicates the maximum spin-down values searched in the analysis.}
\label{fig:ffdotdist}
\end{figure*}

In Fig. \ref{fig:ul_full} the upper limits on signal strain as a function of the frequency are plotted. The values that are valid on just a portion of the corresponding 1-Hz band are labeled by a filled circle. 
A rough comparison with other all-sky searches result shows that our upper limits are better than those established, for instance, by the Einstein@Home pipeline on LIGO S5 data \cite{ref:ein@ho} for frequencies below $\sim$80 Hz (however, in \cite{ref:ein@ho} upper limits are computed every 0.5 Hz, instead of 1 Hz). A useful plot to understand the astrophysical reach of the search is shown in Fig. \ref{fig:ffdotdist}. 
The various sets of points give the relation between the signal frequency derivative and the signal frequency for sources detectable at various distances, assuming their spin-down to be due 
solely to the emission of gravitational radiation (these types of sources are also referred to as {\it gravitars} \cite{ref:gravi}). For instance, considering a source at a distance of 100 
parsecs and emitting a CW signal with a frequency of 80Hz, it would be detectable if its frequency derivative was larger, in modulus, than about $2.8\times 10^{-12}$ Hz/s. On the same plot 
curves of constant ellipticity are also shown. The above source example would have an ellipticity larger than about $2\times 10^{-5}$.    
From this plot we can conclude that our search would have been sensitive enough to detect all the gravitars within about 400 parsecs, emitting a signal with frequency above about 60 Hz and with spin-down age larger than about 4500 years. Also, all the gravitars within about 50 parsecs, emitting a signal with frequency above 20 Hz and with spin-down age larger than about 1500 years would have been detected.  

\section{Conclusions}
\label{sec:concl}
In this paper we have described a low-frequency all-sky search for CW signals, performed using data from the Virgo VSR2 and VSR4 runs. This is the first all-sky search for isolated stars with GW frequencies below 50 Hz. We have applied a hierarchical procedure, in which the incoherent step is based on the FrequencyHough transform. Particular care has been placed on cleaning the data of known and unknown instrumental disturbances, before selecting potentially interesting CW candidates. The criteria used to select candidates have been designed in such a way to avoid sensitivity to any single noise disturbance. For each candidate surviving a coincidence step, a follow-up procedure has been applied in order to reject or confirm it. Three candidates have been found to be most significant, but their potential GW origin has been ruled out after a deeper analysis.
Having found no evidence for CW signals, a 90$\%$ confidence level upper limit on signal strain has been computed over the full frequency band. Our upper limit improves upon prior results below $\sim$80 Hz and the astrophysical reach of the search is interesting. Our search would have been able to detect all the {\it gravitars} within about 400 parsecs from the Earth, spinning at frequencies above 30 Hz and with spin-down age larger than about 4500 years. It would have been also able to detect all the {\it gravitars} within about 50 parsecs from the Earth, spinning at frequencies above 10 Hz and with spin-down age larger than about 1500 years.  
This analysis pipeline will be applied to data of Advanced LIGO and Virgo detectors. Given their improved sensitivities, the chance to detect CW signals by spinning neutron stars will be significantly better. Detection of such signals would provide insights into the internal structure, history formation and demography of neutron stars.    

\appendix

\section{Candidate list}
\label{sec:finalcandlist}
In the following table the main parameters of the final 108 candidates are given. They are ordered according to decreasing rank values. 
\begingroup
\begin{longtable*}{|c|c|c|c|c|c|c|}
\caption{Main parameters of the 108 selected candidates: frequency, spin-down, sky position (in ecliptic coordinates), rank value and the final significance, after doing the follow-up. The candidates are ordered according to their rank value, starting from the most significant.} \\
\hline
Order number & Frequency [Hz]  &  Spin-down $\times 10^{-11}$ [Hz/s] & $\lambda$ [deg] & $\beta$ [deg] & Rank $\times 10^{-4}$ & Significance  \\ \hline
\endfirsthead

\hline
{cont. from prev. page} \\ \hline
Order number & Frequency [Hz] & Spin-down $\times 10^{-11}$ [Hz/s] & $\lambda$ [deg] & $\beta$ [deg] & Rank $\times 10^{-4}$ & Significance  \\
\hline
\endhead

\hline
{cont. on next page} \\ \hline
\endfoot

\hline \hline
\endlastfoot

  1 & 52.8083 & 0 & 276.4196 & -61.0763 & 0.00070508 & $<2.45\cdot 10^{-3}$ \\

  2 & 108.8571 & -0.25027 & 193.2446 & 31.0517 & 0.0014546 & $<2.45\cdot 10^{-3}$\\

  3 & 116.3803 & -11.2223 & 0.1046 & 37.7856 & 0.002162 &  0.18293 \\

  4 & 77.7002 & 0.50253 &  59.7334 & 20.9382 & 0.0025694 &  0.20111 \\

  5 & 103.3949 & 1.9525 & 190.4894 & -25.2949 & 0.0030002 & 0.20243 \\

  6 & 26.221 & 3.2753 & 191.0061 & -65.0493 & 0.0031197 & 0.35941 \\

  7 & 121.8131 & 1.1342 & 119.9904 & 12.8029 & 0.003349 & 0.56723 \\

  8 & 118.8013 & -2.02 & 309.4421 & 60.7465 & 0.0036634 & 0.51951 \\

  9 & 115.2647 & -4.9835 & 132.2512 & 34.3769 & 0.0037117 & 0.72860 \\

 10 & 97.7444 & -3.3448 & 282.8077 & 54.7593 & 0.0071799 & 0.29584 \\

 11 & 100.9588 & -2.5206 & 235.1333 & -13.5452 & 0.0083758 & 0.81853  \\

 12 & 32.6686 & -6.8108 & 341.2171 & 41.1438 & 0.0084554 &  0.15158 \\

 13 & 110.8583 & -5.484 & 332.6659 & -37.837 & 0.008547 & 0.97310 \\

 14 & 55.8628 & -8.2608 & 129.375 & -51.6569 & 0.0093692 & 0.88753 \\

 15 & 123.3888 & -2.9019 & 208.9888 & -31.6621 & 0.009537 & 0.50855 \\

 16 & 68.9449 & 0.57005 & 186.5033 & 41.9996 & 0.0097457 & 0.09512 \\

 17 & 56.0031 & -2.5206 & 335.2227 & -38.9303 & 0.010498 & 0.21112 \\

 18 & 87.4609 & -4.3538 & 252.5236 & 55.405 & 0.011137 & 0.22346 \\

 19 & 125.4042 & 1.7658 & 121.5085 & 18.4552 & 0.011196 & 0.65921 \\

 20 & 33.0337 & -10.4675 & 31.4954 & -49.4705 & 0.012961 & 0.55307 \\

 21 & 124.9926 & -2.7093 & 240.8653 & -20.0556 & 0.013249 & 0.31822 \\

 22 & 95.3279 & -3.0965 & 209.6354 & 46.4176 & 0.013873 & 0.29095 \\

 23 & 67.0831 & -8.7653 & 149.2214 & -70.1448 & 0.014011 & 0.52417 \\

 24 & 119.6512 & -7.3829 & 111.728 & -55.7048 & 0.015368 & 0.76410 \\

 25 & 43.2998 & -0.50451 & 238.3516 & 88.2667 & 0.016197 & $<2.45\cdot 10^{-3}$ \\

 26 & 127.8655 & -4.2248 & 135.7367 & 37.1047 & 0.016758 & 0.51219 \\

 27 & 114.0429 & 0.88388 & 270.2045 & 73.8248 & 0.016903 & 0.82943 \\

 28 & 76.2553 & -4.2208 & 228.0401 & 45.2728 & 0.016994 & 0.37804 \\

 29 & 28.5502 & -9.8955 & 108.4932 & -63.7331 & 0.017029 & 0.21229 \\

 30 & 35.6011 & -0.44095 & 221.1306 & 59.8507 & 0.017121 & 0.14525 \\

 31 & 61.1424 & -7.3134 & 327.0508 & 45.8359 & 0.017213 & 0.97660 \\

 32 & 99.9012 & -0.6336 & 212.2641 & -17.2635 & 0.018999 & 0.33918 \\

 33 & 107.2998 & 1.2632 & 159.1845 & -43.6778 & 0.019099 & 0.74695 \\

 34 & 113.7034 & 0.50451 & 178.7906 & -48.3726 & 0.019143 & 0.74301 \\

 35 & 117.7723 & -2.7073 & 80.7946 & 10.7556 & 0.023097 & 0.07017 \\

 36 & 47.851 & -5.1722 & 27.665 & -39.351 & 0.024202 & 0.46368 \\

 37 & 82.1015 & 1.2632 & 126.6622 & -49.7879 & 0.02424 & 0.12849 \\

 38 & 86.6052 & -10.0882 & 177.278 & 30.7229 & 0.02499 & 0.36871 \\

 39 & 83.0785 & -10.4695 & 342.531 & -32.6498 & 0.025006 & 0.70731 \\

 40 & 65.4443 & -1.0726 & 304.134 & -19.8682 & 0.025727 & 0.16759 \\

 41 & 96.2287 & -7.6888 & 166.1909 & 22.8864 & 0.026087 &  0.67039 \\

 42 & 25.0585 & -2.3378 & 241.7518 & -18.2543 & 0.027399 & 0.47953 \\

 43 & 34.1145 & -5.7998 & 260.6087 & 61.9021 & 0.028246 & 0.88333 \\

 44 & 75.7372 & -10.0901 & 112.5 & 23.0135 & 0.029795 & 0.09669 \\

 45 & 92.1503 & 0.13108 & 172.16 & 63.8249 & 0.032852 & 0.42105 \\

 46 & 51.5143 & 2.024 & 29.8065 & -50.9461 & 0.033757 & 0.10526 \\

 47 & 98.9983 & -6.8724 & 315.3778 & 50.3757 & 0.035169 & 0.268949 \\

 48 & 109.1544 & 2.3954 & 321.2877 & -15.2403 & 0.035476 & 0.29516 \\

 49 & 78.7986 & -2.5206 & 44.8338 & 28.9471 & 0.035727 & 0.08720 \\

 50 & 81.1001 & -3.0906 & 310.0948 & -22.2304 & 0.035921 & 0.37222 \\

 51 & 88.2414 & -4.2267 & 336.0401 & -31.4333 & 0.036207 & 0.27933 \\

 52 & 71.2004 & 1.7678 & 162.6366 & -55.2736 & 0.036235 & 0.01222 \\

 53 & 120.7367 & 0.94744 & 86.9381 & -54.5046 & 0.036489 & 0.82222 \\

 54 & 63.8707 & -4.7273 & 209.8977 & 46.7812 & 0.036588 & 0.44134 \\

 55 & 60.0709 & -7.2538 & 190.8441 & 69.5008 & 0.036653 & 0.24022 \\

 56 & 48.5878 & -3.2852 & 329.1029 & 44.5078 & 0.037649 & 0.42458 \\

 57 & 85.8599 & -4.0361 & 355.2824 & 44.4263 & 0.039416 &  0.27374 \\

 58 & 72.5912 & -9.6472 & 21.228 & 34.9257 & 0.039794 & 0.70244 \\

 59 & 69.598 & -3.2157 & 309.3422 & -24.3772 & 0.039822 & 0.90220 \\

 60 & 91.7067 & -8.1297 & 180.7389 & -35.0032 & 0.040292 & 0.66081 \\

 61 & 23.3888 & -9.3274 & 254.1493 & 59.3324 & 0.040735 & 0.35195 \\

 62 & 41.9977 & -10.1537 & 38.3415 & -58.7027 & 0.041007 & 0.58343 \\

 63 & 106.3632 & -10.6562 & 269.7816 & -9.1791 & 0.041596 & 0.32824 \\

 64 & 112.503 & -6.3719 & 223.3497 & -15.9813 & 0.041966 & 0.03307 \\

 65 & 45.6924 & -9.0176 & 40.9655 & 36.3556 & 0.044055 & 0.97206 \\

 66 & 36.2578 & -9.3294 & 35.7818 & 37.7206 & 0.045444 & 0.93854 \\

 67 & 44.3556 & -10.2788 & 109.623 & 14.9484 & 0.046831 & 0.42458 \\

 68 & 54.0477 & -3.907 & 309.394 & 62.1078 & 0.046967 & 0.53801 \\

 69 & 30.8852 & -1.2593 & 236.7692 & 51.7512 & 0.049419 & 0.01117 \\

 70 & 74.1304 & -0.62964 & 330.9153 & -31.5795 & 0.05038 & 0.59657 \\

 71 & 46.9325 & -1.3864 & 218.8421 & 81.1703 & 0.051493 &  0.40350 \\

 72 & 53.1366 & -5.488 & 342.5228 & -34.7568 & 0.052257 & 0.29516 \\

 73 & 59.0207 & 1.9564 & 248.6587 & -1.5716 & 0.053051 & 0.57457 \\

 74 & 24.0609 & -10.4715 & 108.5902 & -64.411 & 0.053547 & 0.03352 \\

 75 & 122.1853 & -4.4134 & 325.8941 & 48.4439 & 0.057407 & 0.50279 \\

 76 & 94.0253 & 0.44293 & 100.3213 & 10.2366 & 0.057537 & 0.21119 \\

 77 & 104.0544 & -9.4585 & 96.6947 & 9.8744 & 0.057864 & 0.93048 \\

 78 & 105.7558 & -8.8904 & 241.7306 & -31.8533 & 0.058097 & 0.63636 \\

 79 & 80.5196 & -6.0541 & 340.1432 & 45.3848 & 0.060008 & 0.21229 \\

 80 & 21.3803 & -10.3404 & 324.3529 & -34.1692 & 0.062657 & 0.22905 \\

 81 & 66.6324 & -0.25423 & 124.8241 & -57.8314 & 0.062913 & 0.29584 \\

 82 & 102.2397 & -6.4275 & 358.6386 & 47.3049 & 0.063841 & 0.42458 \\

 83 & 126.5903 & -10.8429 & 188.3702 & 51.0238 & 0.065495 & 0.05022 \\

 84 & 90.2319 & -3.0906 & 324.5108 & -24.7639 & 0.066156 & 0.69273 \\

 85 & 101.9708 & -2.2703 & 19.0030 & -61.613 & 0.06835 & 0.92178 \\

 86 & 73.5608 & -0.50451 & 286.4738 & -12.1638 & 0.069204 & 0.61124 \\

 87 & 64.0628 & -4.475 & 246.9913 & -16.5005 & 0.069771 & 0.81564 \\

 88 & 27.3621 & -3.2157 & 228.9322 & 48.5627 & 0.070012 & 0.39766 \\

 89 & 58.5039 & -9.7108 & 131.6129 & -55.3565 & 0.070837 & 0.67039 \\

 90 & 84.5958 & -2.3338 & 103.5869 & -51.7004 & 0.072562 & 0.32960 \\

 91 & 20.2353 & -3.9725 & 87.1304 & -55.6842 & 0.075261 & 0.42222 \\

 92 & 37.7963 & -4.0976 & 178.6723 & 44.0613 & 0.078217 & 0.30726 \\

 93 & 31.032 & -4.6657 & 342.9863 & -35.3718 & 0.081675 & 0.02545\\

 94 & 29.902 & -8.0741 & 243.7377 & 69.6141 & 0.086694 & 0.53216 \\

 95 & 111.1327 & -6.1137 & 23.9462 & 27.6296 & 0.087627 & 0.89821 \\

 96 & 39.6152 & -3.5931 & 229.8023 & 42.5234 & 0.087633 & 0.41899 \\

 97 & 57.1775 & -8.952 & 22.4698 & 25.4575 & 0.095732 & 0.32163 \\

 98 & 70.0715 & -7.7543 & 231.3566 & 43.4535 & 0.097824 & 0.46943 \\

 99 & 42.9996 & -0.31977 & 72.6974 & 14.6357 & 0.098683 & 0.25917 \\

100 & 49.1547 & 0.065538 & 286.3941 & -14.3416 & 0.10996 & 0.67597 \\

101 & 62.2854 & -6.5625 & 155.5203 & -86.2875 & 0.11363 & 0.07262 \\

102 & 38.8918 & -7.3769 & 229.5254 & -17.7099 & 0.11676 & 0.44134 \\

103 & 93.0512 & -3.3409 & 297.0175 & -21.1575 & 0.12164 & 0.40782 \\

104 & 40.4971 & -1.3268 & 123.7959 & 31.1944 & 0.13931 & 0.48603 \\

105 & 79.9665 & -11.3455 & 230.6667 & -10.579 & 0.15052 & 0.46368 \\

106 & 22.5768 & -9.391 & 357.9813 & 33.7066 & 0.16427 & 0.38596 \\

107 & 89.1945 & -9.0811 & 278.9743 & 51.5108 & 0.16766 & 0.23463 \\

108 & 50.8034 & 0.37739 & 84.3333 & -40.3263 & 0.26302 & 0.92441 \\
\hline
\label{tab:SELb}
\end{longtable*}
\endgroup

\section{Outliers analysis}
\label{sec:outl}
In this section we describe the further analysis steps applied to the three outliers found in the follow-up step (see Sec. \ref{sec:followup}). The goal here is to reject or confirm each of them as potential gravitational wave signal candidates. The main parameters of the three outliers are listed in Tab. \ref{tab:outliers}. 
\begin{table}[!htbp]
\begin{center}
\begin{tabular}{|c|c|c|c|c|}
\hline
Significance  & $f$ [Hz]  & $\dot{f}$ [Hz/s] & $\lambda$ [deg] & $\beta$ [deg] \\
\hline
0.0114 & 30.88541 & $-9.19\times 10^{-12}$ & 235.40 & 50.99   \\
\hline
$<$0.0024 & 43.29974 & $-5.04\times 10^{-12}$ & 198.30 & 89.42  \\
\hline
0.0112 & 71.20035 & $1.13\times 10^{-11}$ & 162.83 & -55.27 \\
\hline
\end{tabular}
\caption{Main parameters of the three outliers found in the follow-up. The significance is given in the second column; $f$ and $\dot{f}$ are the candidate frequency and spin-down; $\lambda$ and $\beta$ is the candidate position in ecliptic coordinates. They correspond, respectively, to candidates number 69, 25 and 52 in Tab. V. Note that the reported parameter values are those after the follow-up step and then are slightly different from those in Tab. V which, on the contrary, have been computed before the follow-up.} 
\label{tab:outliers}
\end{center}
\end{table}
\subsection{Outlier at 30.88 Hz}
\begin{figure*}[!htbp]
\includegraphics[width=12cm]{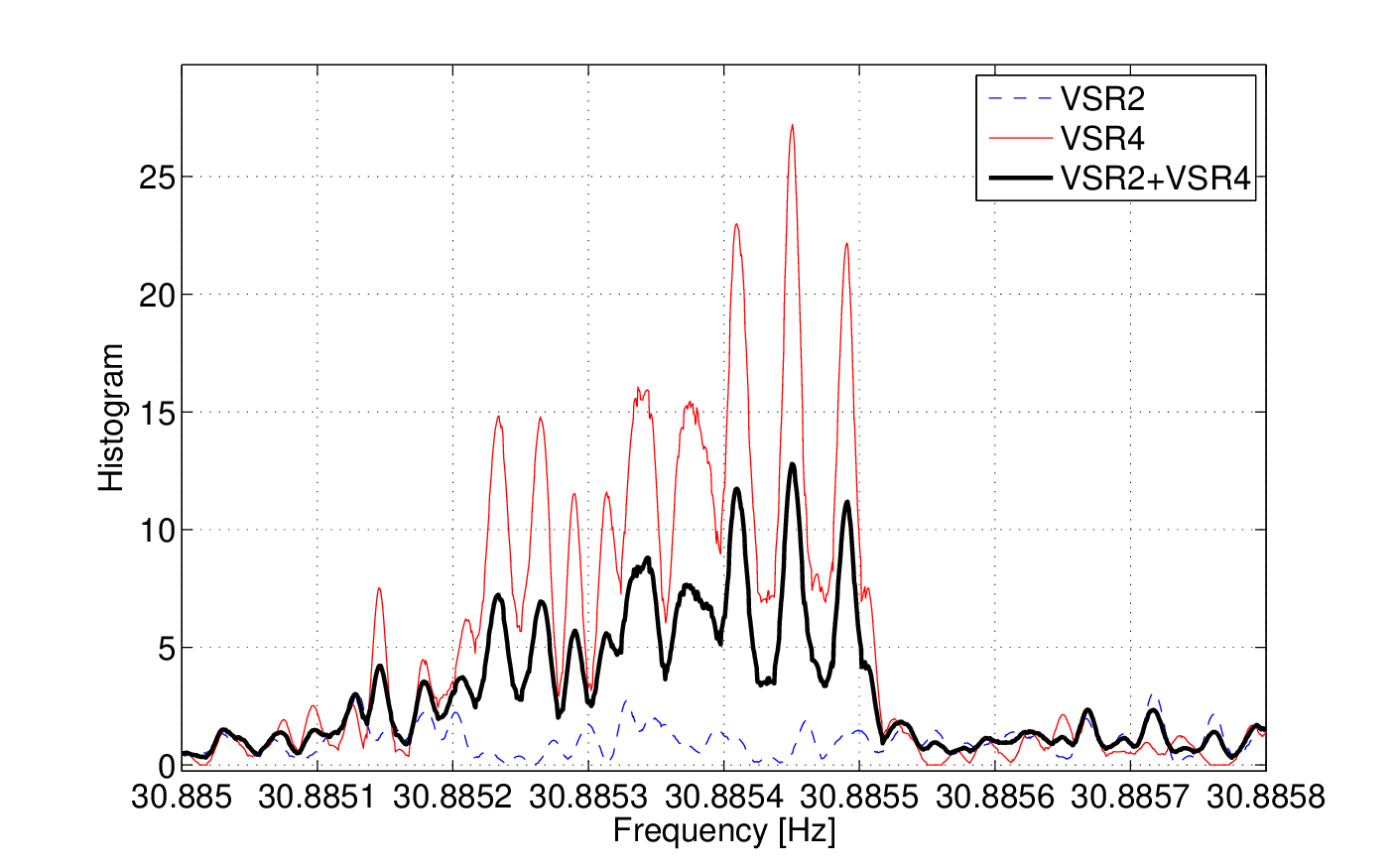} 
\caption{Final peakmap histograms for the outliers at 30.88 Hz. The dark thick curve corresponds to the full VSR2/VSR4 analysis; the thin dashed (blue in color version) curve corresponds to VSR2 analysis while the thin continuous (red in color version) curve corresponds to the analysis of VSR4 alone.}
\label{fig:30_cand_runs_p}
\end{figure*}

\begin{figure*}[!htbp]
\includegraphics[width=8cm]{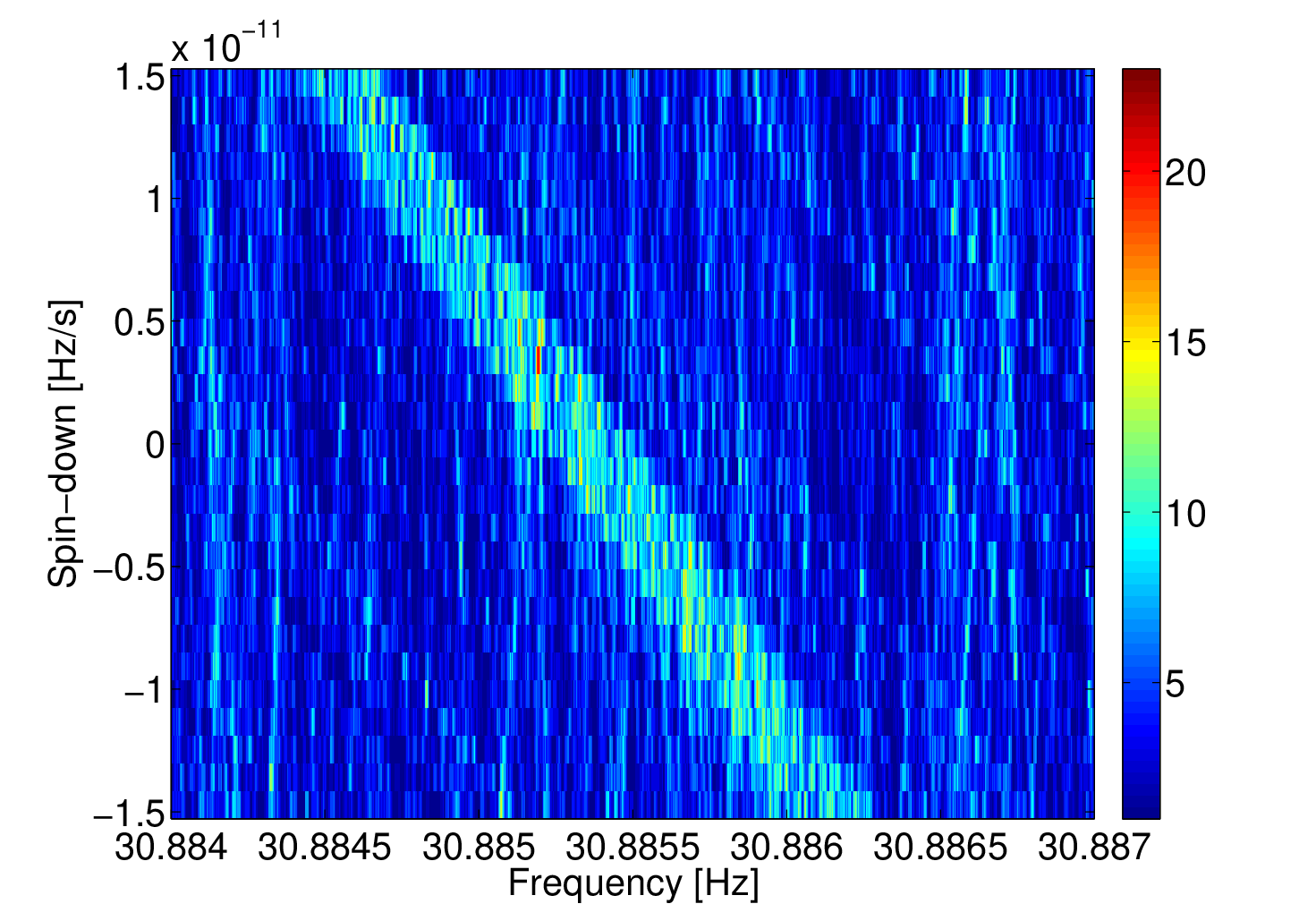} 
\includegraphics[width=8cm] {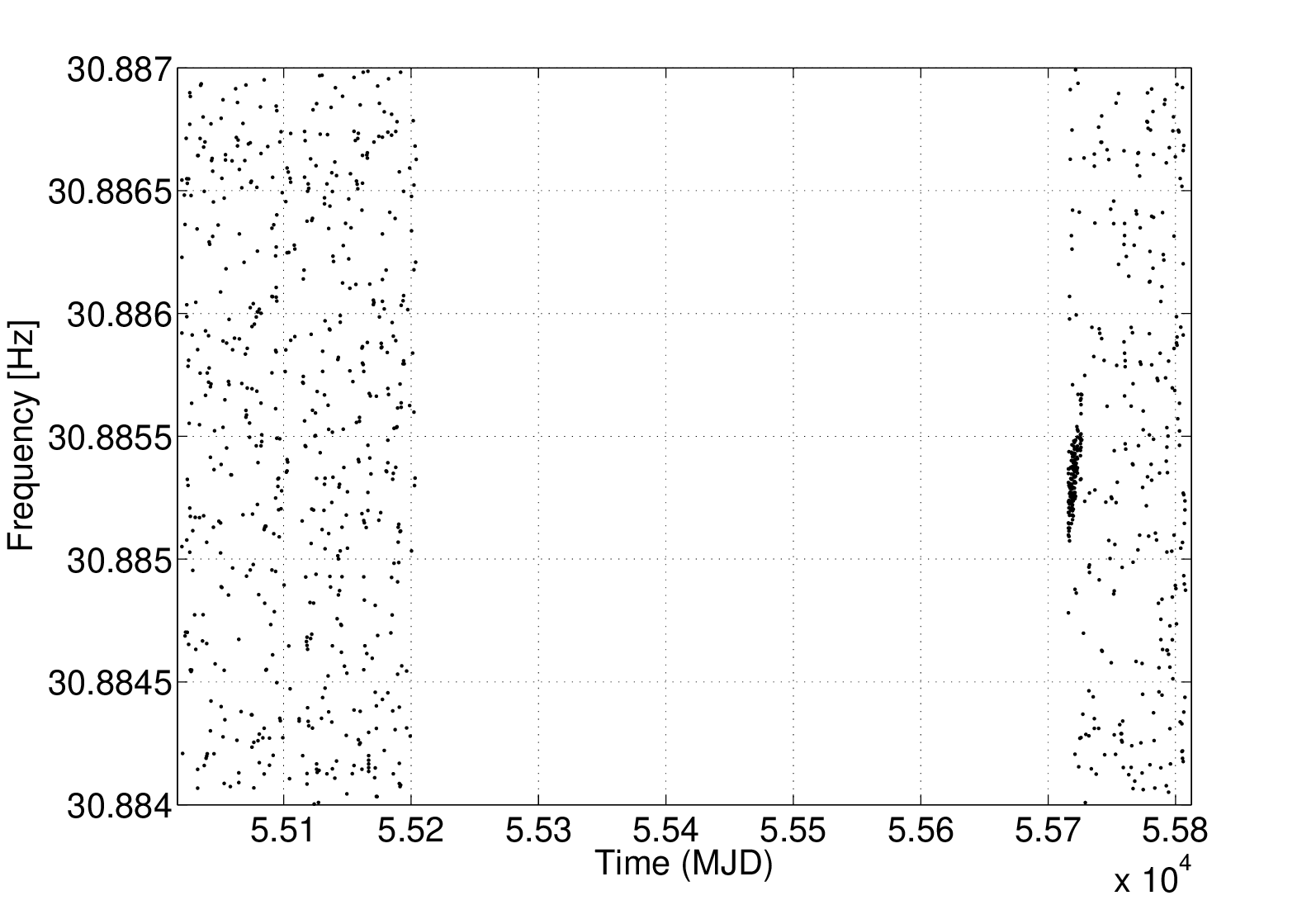}
\caption{Hough map (left) and peakmap (right) for the outlier at 30.88 Hz. The right plot shows the combination of the peakmaps from VSR2 and VSR4 data. The empty region in the middle corresponds to the time separation between the end of VSR2 and the beginning of VSR4. The outlier at 30.88 Hz is due to the concentration of peaks clearly visibile at the beginning of VSR4.}
\label{fig:30_cand_hmpm_p}
\end{figure*}

This candidate has a significance just above the 1$\%$ threshold. An important verification step consists in checking if the ``signal characteristics'' are consistent among the two detectors. For this purpose we have repeated the last part of the follow-up procedure (from the FrequencyHough stage on) separately for the two runs. The result is shown in Fig. \ref{fig:30_cand_runs_p}, where the final peakmap histograms are plotted both for the two runs separately and for the joint analysis. We clearly see that the outlier is strong in VSR4 data but not in VSR2 data. Given the characteristics of the two runs this is not what we expect for a real GW signal. The expected ratio between the peakmap histogram
heights for the two data runs can be estimated for nominal signal strengths \cite{ref:freqhoughmethod} and is about 1.5. This is much lower than the observed ratio, which is about 13. The inconsistency of the apparent signals in VSR2 and VSR4 is also indicated by the lower height of the joint-run peak map
histogram in Fig.~\ref{fig:30_cand_runs_p} with respect to that of VSR4 alone. 
This is confirmed by the Hough map, shown in Fig. \ref{fig:30_cand_hmpm_p} (left), where a single and wide stripe is clearly visible, suggesting that the outlier is produced by a rather short duration disturbance present in VSR4 data only (for comparison, Fig. 3 in \cite{ref:freqhoughmethod} shows the Hough map for an HI in VSR2 data). 
This seems to be confirmed looking at the peakmap around the candidate [see Fig. \ref{fig:30_cand_hmpm_p} (right)], where a high concentration of peaks, of unidentified origin, is present at the beginning of VSR4, lasting for about 10 days. From this analysis we think the outlier can be safely ruled out as being inconsistent with a real GW signal.       

\subsection{Outlier at 43.30 Hz}
\begin{figure*}[!htbp]
\includegraphics[width=12cm]{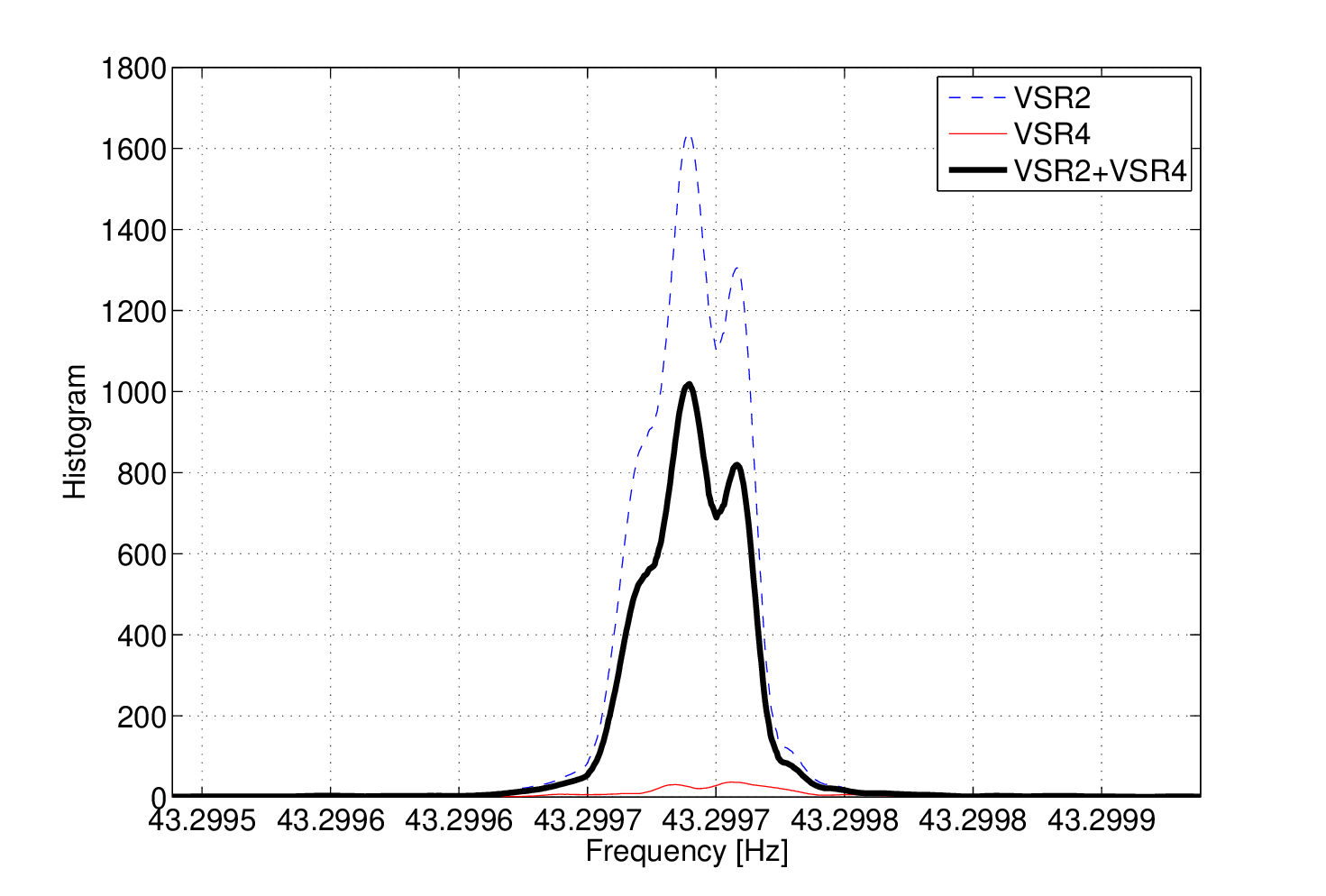} 
\caption{Final peakmap histograms for the outliers at 43.30 Hz. The dark thick curve corresponds to the full VSR2/VSR4 analysis; the light continuous (red in color version) curve corresponds to VSR4 
analysis while the light dashed (blue in color version) curve corresponds to the analysis of VSR2 alone.}
\label{fig:43_cand_runs_p}
\end{figure*}

\begin{figure*}[!htbp]
\includegraphics[width=8cm]{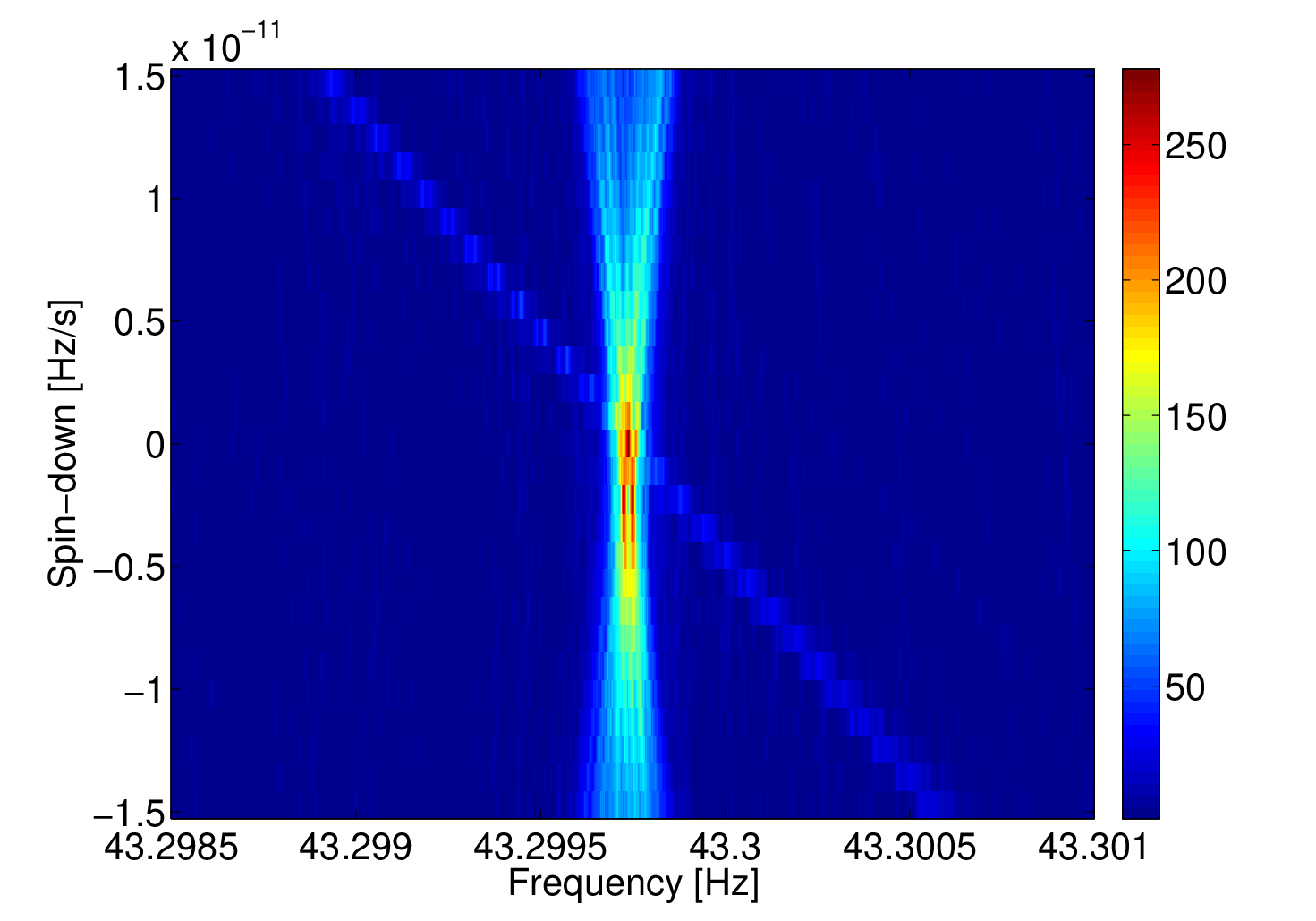} 
\includegraphics[width=8cm] {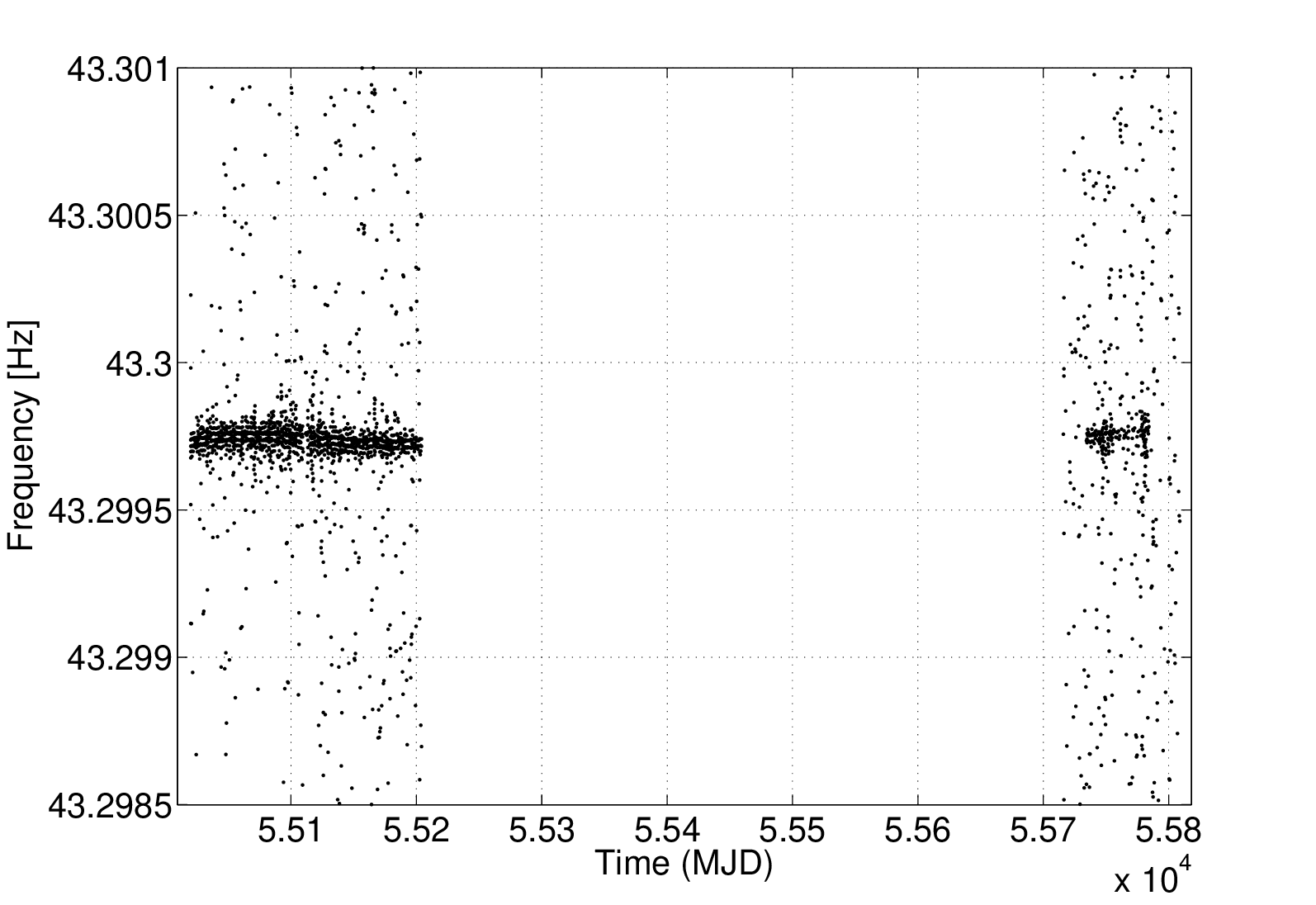}
\caption{Hough map (left) and peakmap (right) for the outlier at 43.30 Hz.}
\label{fig:43_cand_hmpm_p}
\end{figure*}

This candidate has a very high significance (corresponding to the smallest possible p-value). In  Fig. \ref{fig:43_cand_runs_p} the peakmap histogram for the full analysis, together to those for the single runs are shown. 
We see that basically the ``signal'' is present mainly in one of the two runs. 
This is also what we conclude looking at the Hough map and the peakmap in Fig. \ref{fig:43_cand_hmpm_p}, 
where the main contribution coming from VSR2 is clearly visible.  
From the previous considerations it seems that also this outlier is associated to some disturbance in the data. In particular it happens in a frequency region (between 40 and 45 Hz) which 
is heavily polluted by noise produced by rack cooling fans. It must be noticed that outliers clearly associated to a disturbed region found at this step of the procedure should not be surprising. In fact, despite
all the cleaning procedures and criteria used to select candidates, some instrumental disturbance can be still present in the data. This is the reason why the most interesting candidates must be subject to a close scrutiny.

\subsection{Outlier at 71.20 Hz}
\begin{figure*}[!htbp]
\includegraphics[width=12cm]{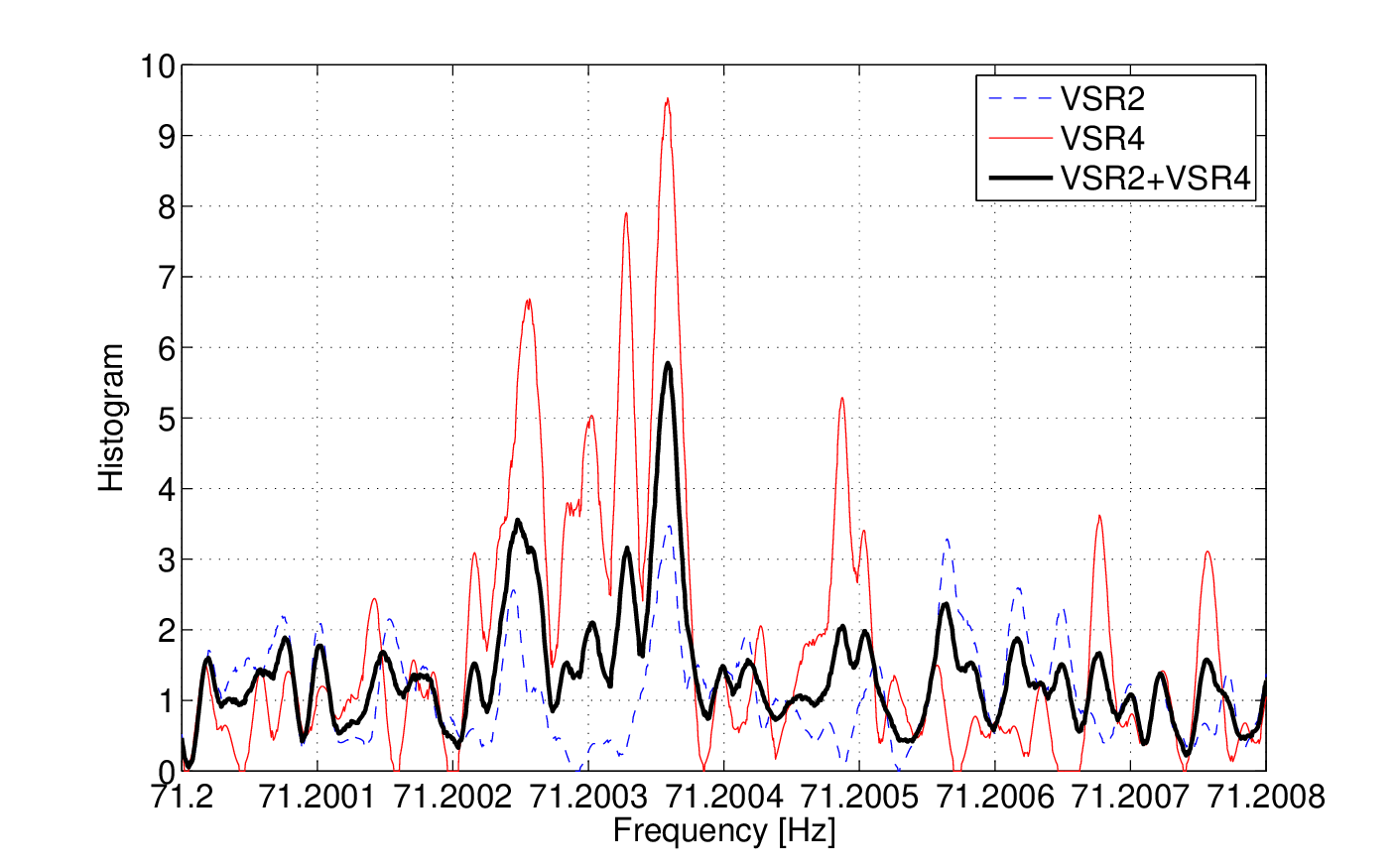} 
\caption{Final peakmap histograms for the outlier at 71.20 Hz. The dark thick curve corresponds to the full VSR2/VSR4 analysis; the light dashed (blue in color version) curve corresponds to VSR2 
analysis while the light continuous (red in color version) curve corresponds to the analysis of VSR4 alone.}
\label{fig:71_cand_runs_p}
\end{figure*}

\begin{figure*}[!htbp]
\includegraphics[width=8cm]{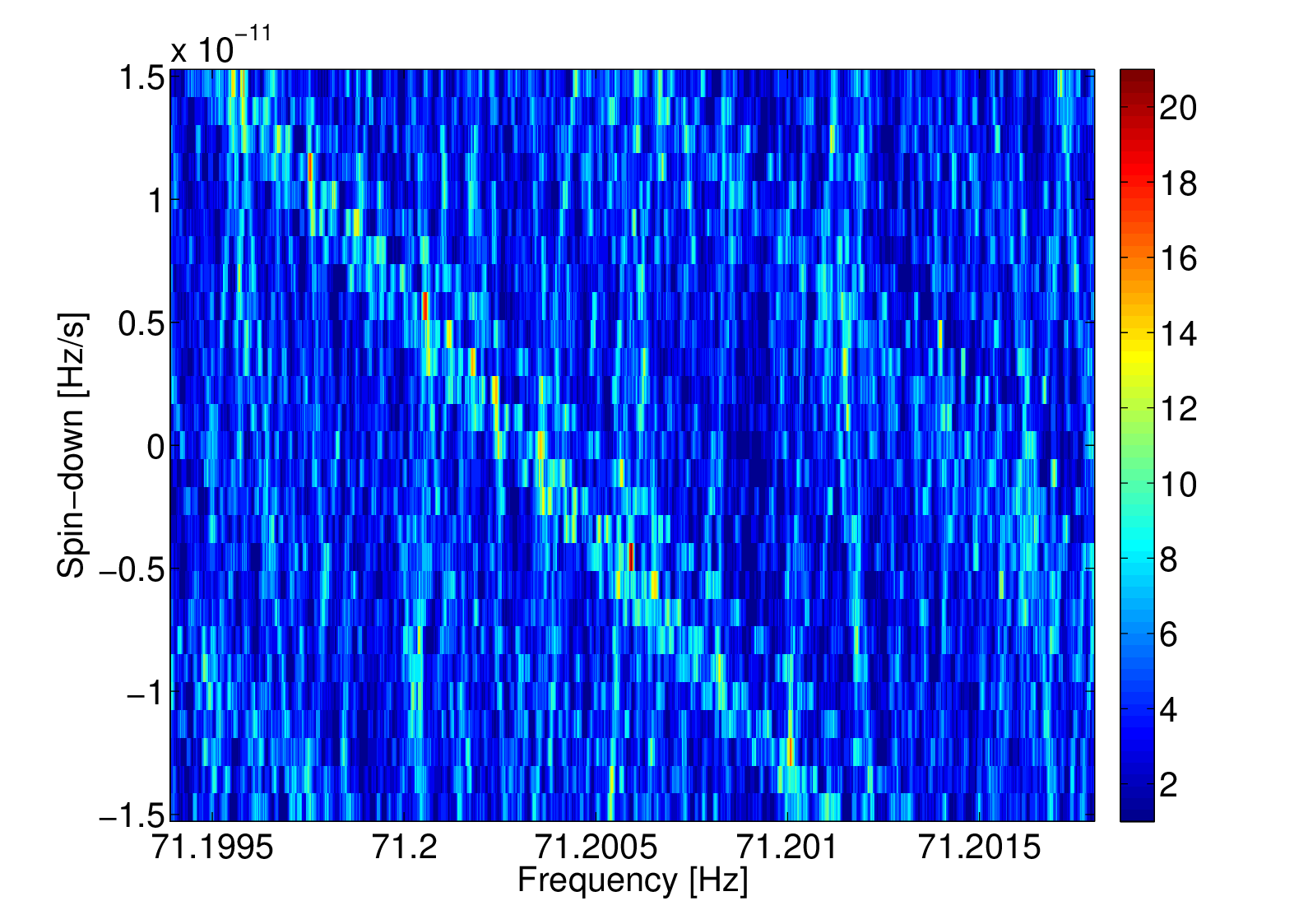} 
\includegraphics[width=8cm] {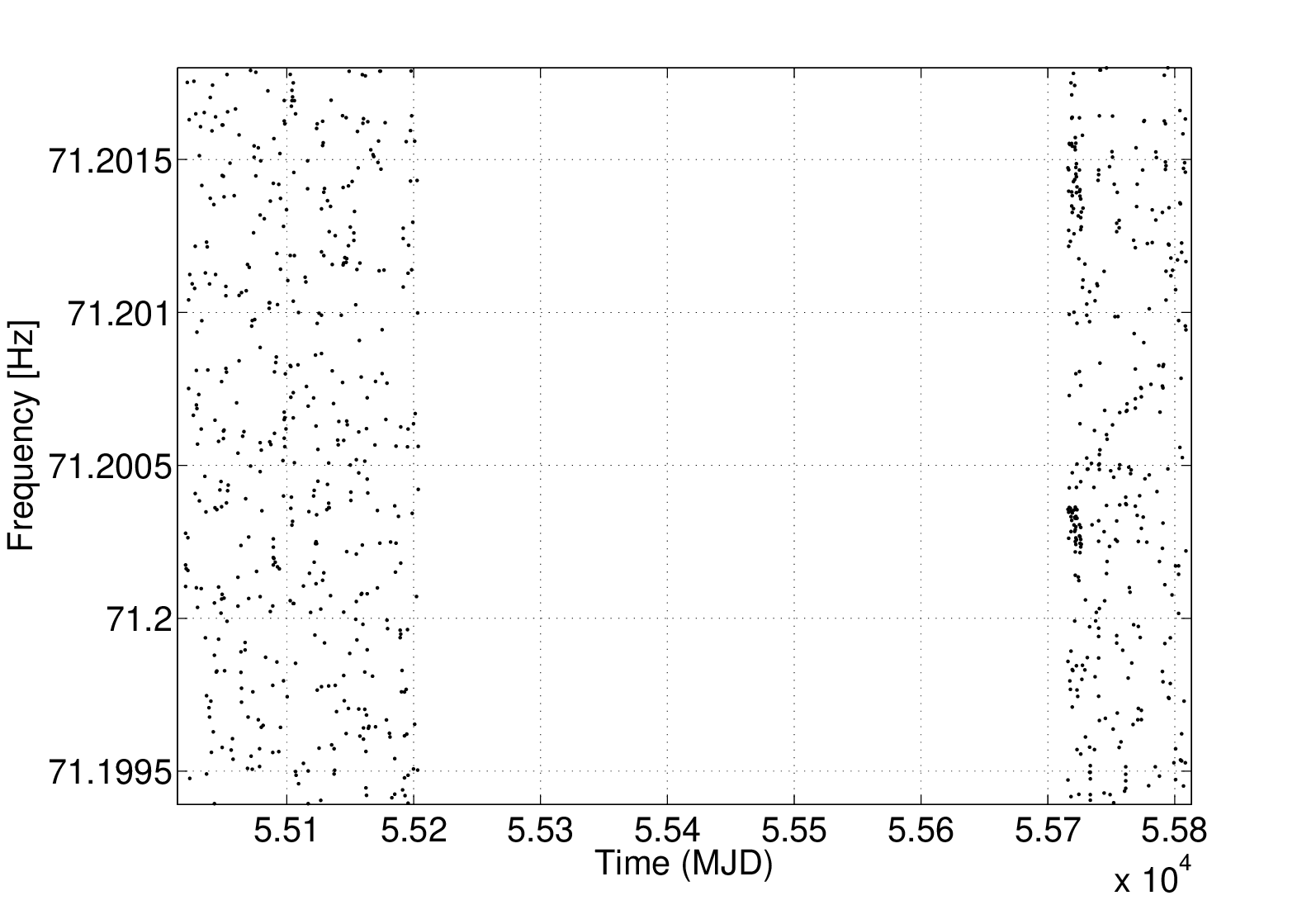}
\caption{Hough map (left) and peakmap (right) for the outlier at 71.20 Hz. As for the outlier at 30.88 Hz, this outlier is due to a concentration of peaks at the beginning of VSR4, barely visible in the right plot.}
\label{fig:71_cand_hmpm_p}
\end{figure*}

This third outlier has many characteristics in common with that at 30.88 Hz. Comparing the peakmap histograms of Figs. \ref{fig:30_cand_runs_p} and \ref{fig:71_cand_runs_p}, we see again that it is mainly present in VSR4 data. In this case the expected ratio between the peakmap histogram
heights, in presence of a signal, is about 0.7 while we observe a ratio of about 3. Moreover, as for the outlier at 30.88 Hz, the height of the joint-run peak map histogram is smaller than for VSR4 alone, which is also inconsistent with the signal hypothesis. 
This is confirmed by the Hough map shown in Fig. \ref{fig:71_cand_hmpm_p} (left), where basically only a wide stripe in the frequency/spin-down plane is visibile, indicating that the 
outlier comes mainly from only one of the two runs. A further confirmation comes from the peakmap in Fig. \ref{fig:71_cand_hmpm_p} (right), in which a high concentration of peaks is visible at the right 
frequencies just at the beginning of VSR4, similarly to what happens for the first outlier.  
We conclude that this outlier is not due to a real GW signal.
\newpage
\section{Excluded bands}
\label{sec:excluded}
In the following table the noisy frequency intervals excluded from the computation of the upper limits are given. The two intervals 52-53 Hz and 108-109 Hz have been excluded because their most significant candidate is an hardware injected signal. 
\newpage 
\begin{table*}[!htbp]
\begin{center}
\begin{tabular}{|c|c|}
\hline
Band [Hz] & Excluded range [Hz] \\ \hline \hline
20 - 21 & [20,20.17505] \\ \hline
22 - 23 & [22.34997,22.43506] \\ \hline
23 - 24 & [23.68359,23.77637] \\ \hline
25 - 26 & [25.10864,25.15381], [25.15967,25.24634] \\ \hline
26 - 27 & [26.41052,26.43799] \\ \hline
28 - 29 & [28.78516,28.82019] \\ \hline
29 - 30 & [29.08484,29.09985], [29.10010,29.13440] \\ \hline
30 - 31 & [30.82519,30.85510] \\ \hline
34 - 35 & [34.54517,34.58508] \\ \hline
35 - 36 & [34.99304,35.05713], [35.27398,35.30729], [35.63635,35.78503] \\ \hline
\multirow{2}{*}{37 - 38} & [37.05517,37.15515], [37.27429,37.31531], [37.38379,37.39734] \\
& [37.50513,37.62573], [37.70520,37.77563], [37.96509,37.99988] \\ \hline
38 - 39 & [38,38.02795], [38.02820,38.04565], [38.16516,38.20740] \\ \hline
39 - 40 & [39.95520,39.99988] \\ \hline
40 - 41 & [40,40.07568] \\ \hline
41 - 42 & [41.10522,41.13501], [41.50329,41.53833], [41.92456,41.98804] \\ \hline
42 - 43 & [42.01514,42.04700], [42.67517,42.76758], [42.80493,42.85730] \\ \hline
\multirow{2}{*}{43 - 44} & [43.22302,43.30041], [43.30359,43.33606], [43.69189,43.71558] \\
& [43.85803,43.93787] \\ \hline
45 - 46 & [45.96863,45.99939] \\ \hline 
\multirow{2}{*}{46 - 47} & [46.00024,46.08508], [46.29724,46.31506], [46.36523,46.41504] \\
& [46.60522,46.69507], [46.81738, 46.89501] \\ \hline
47 - 48 & [47,47.02551], [47.06518,47.10510] \\ \hline
48 - 49 & [48.83484,48.87512] \\ \hline
\multirow{2}{*}{49 - 50} & [49.38806,49.403564], [49.40588,49.62561], [49.67956,49.75683] \\
& [49.83520,49.95080], [49.95105,49.99890] \\ \hline
\multirow{2}{*}{50 - 51} & [50,50.05041], [50.05066,50.07971], [50.07995,50.27893] \\
& [50.29407,50.60705] \\ \hline
\multirow{2}{*}{51 - 52} & [51.01477,51.07690], [51.18506,51.22534], [51.38513,51.41504] \\
& [51.59521,51.63525] \\ \hline
52 - 53 & [52.0,53.0] \\ \hline
56 - 57 & [56.13452,56.15942] \\ \hline
60 - 61 & [60.25024,60.27502], [60.81225,60.82788] \\ \hline
\multirow{3}{*}{61 - 62} & [61.32324,61.35095], [61.47375,61.49219], [61.50574,61.51977] \\
& [61.52026,61.58581], [61.61938,61.63513], [61.67663,61.71472] \\
& [61.71496,61.97717] \\ \hline
\multirow{2}{*}{62 - 63} & [62,62.25500], [62.29517,62.31323], [62.31396,62.32690] \\
& [62.38623,62.47876], [62.48059,62.52502] \\ \hline
63 - 64 & [63.19714,63.22656], [63.75146,63.77502] \\ \hline
82 - 83 & [82.22692,82.24511] \\ \hline
84 - 85 & [84.81482,84.89502] \\ \hline
90 - 91 & [90.49523,90.52685] \\ \hline
\multirow{3}{*}{92 - 93} & [92.00756,92.02527], [92.02746,92.07629], [92.264523,92.27991] \\
& [92.30432,92.32837], [92.32861,92.36609], [92.39587,92.47815] \\
& [92.49524,92.73511], [92.93469,92.97558] \\ \hline
99 - 100 & [99.92712,99.97461] \\ \hline
100 - 101 & [100,100.04614], [100.21228,100.23535] \\ \hline
102 - 103 & [102.78369,102.80505] \\ \hline
108 - 109 & [108.0 109.0] \\ \hline
111 - 112 & [111.76636,111.80127], [111.81116,111.87512] \\ \hline
\multirow{2}{*}{112 - 113} & [112.47681,112.49194], [112.50500,112.51843], [112.52331,112.59997] \\
& [112.61475,112.65307], [112.67578,112.83032], [112.83374,112.87512] \\ \hline
\multirow{3}{*}{113 - 114} & [113.00012,113.03137], [113.04187,113.08520], [113.11633,113.14355] \\
& [113.17065,113.28479], [113.28552,113.38379], [113.38513,113.47986] \\
& [113.48010,113.52502], [113.59570,113.62610], [113.63488,113.66504] \\ \hline
114 - 115 & [114.11096,114.15332], [114.15515,114.22534] \\ \hline
127 - 128 & [127.98523,127.99841] \\ \hline
\hline
\end{tabular}
\caption{Frequency intervals excluded in the computation of the upper limit. The total vetoed frequency band amounts to about 8.6 Hz.}
\label{tab:exc_band}
\end{center}
\end{table*}

\bibliography{Main}

\section*{Acknowledgement}

The authors gratefully acknowledge the support of 
the United States National Science Foundation (NSF) for the construction and operation of the LIGO Laboratory,
the Science and Technology Facilities Council (STFC) of the United Kingdom, 
the Max-Planck-Society (MPS), and the State of Niedersachsen/Germany 
for support of the construction and operation of the GEO600 detector,
the Italian Istituto Nazionale di Fisica Nucleare (INFN) and 
the French Centre National de la Recherche Scientifique (CNRS)
for the construction and operation of the Virgo detector
and the creation and support  of the EGO consortium. 
The authors also gratefully acknowledge research support from these agencies as well as by 
the Australian Research Council,
the International Science Linkages program of the Commonwealth of Australia,
the Council of Scientific and Industrial Research of India, 
Department of Science and Technology, India,
Science \& Engineering Research Board (SERB), India,
Ministry of Human Resource Development, India,
the Spanish Ministerio de Econom\'ia y Competitividad,
the Conselleria d'Economia i Competitivitat and Conselleria d'Educaci\'o, Cultura i Universitats of the Govern de les Illes Balears,
the Foundation for Fundamental Research on Matter supported by the Netherlands Organisation for Scientific Research, 
the National Science Centre of Poland, 
the European Union,
the Royal Society, 
the Scottish Funding Council, 
the Scottish Universities Physics Alliance, 
the National Aeronautics and Space Administration, 
the Hungarian Scientific Research Fund (OTKA),
the Lyon Institute of Origins (LIO),
the National Research Foundation of Korea,
Industry Canada and the Province of Ontario through the Ministry of Economic Development and Innovation, 
the National Science and Engineering Research Council Canada,
the Brazilian Ministry of Science, Technology, and Innovation,
the Carnegie Trust, 
the Leverhulme Trust, 
the David and Lucile Packard Foundation, 
the Research Corporation, 
and the Alfred P. Sloan Foundation.
The authors gratefully acknowledge the support of the NSF, STFC, MPS, INFN, CNRS and the
State of Niedersachsen/Germany for provision of computational resources. 
The authors are also grateful to the anonymous referees for their comments, which helped to improve the clarity of the paper. 
\end{document}